\documentclass[12pt]{article}

\PassOptionsToPackage{round,authoryear}{natbib}

\usepackage{arxiv}

\usepackage[utf8]{inputenc}
\usepackage{microtype}
\usepackage{amsmath,amssymb}
\usepackage{mathtools}

\allowdisplaybreaks

\usepackage{xfrac}
\usepackage{xcolor}
\usepackage{comment}
\usepackage{graphicx}
\usepackage{tikz}
\usepackage{tikz-cd}
\usetikzlibrary{patterns}
\usepackage{booktabs}
\usepackage{multirow}

\usepackage{caption}
\usepackage{enumitem}

 \usetikzlibrary{positioning, shapes.geometric, arrows.meta, fit, decorations.pathreplacing, calc}
\usepackage{algorithm}
\usepackage{algorithmic}

\usepackage{natbib}
\usepackage{amsthm}
\theoremstyle{plain}
\newtheorem{theorem}{Theorem}
\newtheorem{proposition}{Proposition}
\newtheorem{lemma}{Lemma}
\theoremstyle{definition}
\newtheorem{definition}{Definition}
\newtheorem{assumption}{Assumption}
\theoremstyle{remark}
\newtheorem{remark}{Remark}
\usepackage{url}

\graphicspath{{./art/}}

\usepackage{authblk}

\providecommand{\PP}{\mathbb{P}}
\providecommand{\EE}{\mathbb{E}}
\providecommand{\RR}{\mathbb{R}}
\providecommand{\ind}{\mathbf{1}}
\providecommand{\one}{\mathbf{1}}
\providecommand{\rd}{\mathrm{d}}
\providecommand{\cC}{\mathcal{C}}
\providecommand{\cY}{\mathcal{Y}}
\providecommand{\cN}{\mathcal{N}}
\providecommand{\cA}{\mathcal{A}}

\providecommand{\pb}{\mathbf{p}}
\providecommand{\eb}{\mathbf{e}}
\providecommand{\iid}{\stackrel{\text{i.i.d.}}{\sim}}
\providecommand{\cI}{\mathcal{I}}
\providecommand{\cM}{\mathcal{M}}
\providecommand{\cL}{\mathcal{L}}

\providecommand{\Zb}{\mathbf{Z}}
\providecommand{\bmu}{\boldsymbol{\mu}}

\providecommand{\wvert}{\,|\,}
\providecommand{\texorpdfstring}[2]{#2}
\makeatletter
\@ifundefined{argmin}{}{}
\@ifundefined{argmax}{}{}
\@ifundefined{Var}{\DeclareMathOperator{\Var}{Var}}{}
\makeatother
\DeclareMathOperator{\Unif}{Unif}

\title{Data-light Uncertainty Set Merging with Admissibility}

	\author[,1]{\fontsize{13pt}{12pt}\selectfont Shenghao Qin\footnote{Authors contributed equally.}}
	\author[$*$,2]{Jianliang He}
	\author[$*$,1]{Qi Kuang}
	\author[1]{Bowen Gang}
	\author[1]{Yin Xia}

	\affil[1]{Department of Statistics and Data Science, Fudan University}
	\affil[2]{Department of Statistics and Data Science, Yale University}

\begin{document}
\maketitle

\begin{abstract}
	This article introduces a Synthetics, Aggregation, and Test inversion (SAT) approach for merging diverse and potentially dependent uncertainty sets into a single unified set. The procedure is data-light, relying only on initial sets and their nominal levels, and it flexibly adapts to user-specified input sets with possibly varying coverage guarantees. SAT is motivated by the challenge of integrating uncertainty sets when only the initial sets and their control levels are available—for example, when merging confidence sets from distributed sites under communication constraints or combining conformal prediction sets generated by different algorithms or data splits. To address this, SAT constructs and aggregates novel synthetic test statistics, and then derive merged sets through test inversion. Our method leverages the duality between set estimation and hypothesis testing, ensuring reliable coverage in dependent scenarios. A key theoretical contribution is a rigorous analysis of SAT's properties, including  its admissibility in the context of deterministic set merging. Both theoretical analyses and empirical results confirm the method’s finite-sample coverage validity and desirable set sizes. 
  		
\end{abstract}


\section{Introduction}\label{sec:intro}

Uncertainty sets, such as confidence intervals and prediction intervals, are pivotal in statistical inference as they facilitate accurate representation and management of data variability. The integration of these sets has wide-ranging applications across various fields. However, considerable challenges emerge, especially when only the initial uncertainty sets and their control levels are available, along with possible intrinsic dependencies among the sets. To highlight the significance of set merging, we will first explore two prominent examples.

\medskip

\emph{Example 1 (Distributed Learning with Communication Constraint).}
In distributed learning, the primary objective is to collaboratively make inferences using data distributed across different studies. Recent advancements have focused on distributed mean estimation \citep{cai2024distributed}, prediction \citep{humbert2023one}, and causal inference \citep{xiong2023federated}. A major challenge in distributed learning is the presence of communication constraints, which can arise from bandwidth limitations, privacy concerns, or cost considerations. These constraints restrict the amount of information that can be exchanged between studies. In certain scenarios, local sites can only transmit the confidence set and its associated confidence level to a central aggregator, highlighting the necessity for effective and data-light methods to merge these uncertainty sets for robust and reliable inferences.

\medskip

\emph{Example 2 (Algorithmic Stability and Derandomization).}
Conformal prediction, pioneered by \cite{vovk2005algorithmic}, has gained considerable popularity due to its minimal assumption requirements and its capability to provide finite-sample valid prediction sets for any black-box models. One of the most widely adopted versions is split conformal inference \citep[e.g.,][]{lei2018distribution,angelopoulos2021gentle}, which is valued for its computational efficiency. However, the resulting prediction set can be influenced by the way the data is split. Additionally, variations in the prediction set may occur depending on the algorithm employed to compute the non-conformity score. To mitigate these issues, combining different prediction sets becomes a natural and necessary strategy.

This paper aims to develop an efficient and flexible method to combine $L$ different--potentially dependent--uncertainty sets into a single set, given only the initial uncertainty sets and their corresponding control levels. 
We term this method a \emph{data-light} approach, emphasizing that it requires no access to raw data and the processes used to construct the initial uncertainty sets.
Formally, let $Y$ denote the prediction target, and let $\cC_{\ell}$ represent any initial uncertainty set from the $\ell$-th study such that
\begin{equation}
	\PP(Y\notin \cC_{\ell})\leq\alpha_\ell,\quad \ell=1,\ldots,L,
	\label{eq:miscoverage}
\end{equation}
where $\alpha_\ell$'s are possibly varying control levels. 
The goal is to construct a merged set $\bar \cC_{\alpha}$ such that $\PP(Y\notin\bar \cC_{\alpha})\leq\alpha$ for a pre-specified $\alpha\in(0,1)$, using only the available $\{(\cC_{\ell}, \alpha_\ell)\}_{\ell=1}^L$, while also ensuring that the merged set remains small in size. 

The aggregation of uncertainty sets has gained considerable interest recently, especially in the context of conformal prediction. 
\cite{yang2024selection} proposes selecting from nested conformal prediction sets the one with the smallest size, incorporating either coverage level adjustments or additional sample splitting.
\cite{liang2024conformal} advances this approach by leveraging the properties of full conformal prediction, ensuring that coverage levels are always guaranteed with relatively small set sizes.
Additionally, \cite{stutz2021learning} focuses on training an optimal classifier that generates small conformal prediction sets by evaluating set size using a subset of mini-batch data during gradient descent. 
The main idea behind these methods is to choose an optimal non-conformity score to minimize the prediction set size, rather than merging the resulting sets. 
Furthermore, these approaches implicitly assume that all sets being aggregated have the same coverage guarantee.

In a different line of research, \cite{chen2021learning,bai2022efficient,fan2023utopia,kiyani2024length} propose using constrained optimization approaches to directly minimize set size while maintaining coverage. 
However, these methods require access to the original data, which we do not assume to be available. 
Recently, \cite{cherubin2019majority,solari2022multi,gasparin2024merging} have explored merging uncertainty sets through majority voting. However, the admissibility of the majority voting method is not studied. In later sections, we show that our method can be viewed as a generalization of the voting method. In addition, we provide a theoretical analysis of the proposed approach, establishing key properties including its admissibility.

In this paper, we develop a novel data-light procedure which we term Synthetics, Aggregation, and Test inversion (SAT). The core of the SAT framework is to convert the problem of merging sets into a more tractable statistical problem of aggregating evidence. This is achieved in three steps. First, we construct novel synthetic test statistics, such as e-values or p-values, that depend solely on the initial uncertainty sets and their coverage levels. These statistics are designed to mimic the true, underlying unknown statistics used to construct the initial sets. Second, these synthetic test statistics are merged from the multiple input sets. Finally, the merged uncertainty set is derived through the test inversion of the aggregated synthetic statistic. A schematic overview of the SAT procedure is given in Figure~\ref{fig:sat_schematic_candidate}.

\begin{figure}[ht]
	\centering
	\begin{tikzpicture}[
		node distance=0.8cm and 2.2cm,
		candidate/.style={ellipse, draw, thick, minimum size=1cm, fill=blue!10},
		syn_stat/.style={rectangle, draw, thick, minimum width=1.5cm, minimum height=0.7cm, fill=teal!10},
		agg_stat_pass/.style={rectangle, draw, thick, minimum width=1.5cm, minimum height=0.7cm, fill=green!20},
		agg_stat_fail/.style={rectangle, draw, thick, minimum width=1.5cm, minimum height=0.7cm, fill=red!20},
		final_set/.style={ellipse, draw=red!80, thick, dashed, inner sep=0.4cm, rounded corners=15pt},
		arrow/.style={-Latex, thick},
		brace/.style={decoration={brace,amplitude=10pt},thick}
		]
		
		\node[candidate] (y1) {$y_1$};
		\node[candidate, right=of y1] (y2) {$y_2$};
		\node[right=of y2] (ydots) {$\cdots$};
		\node[candidate, right=of ydots] (ym) {$y_m$};
		
		
		\node[syn_stat, below=of y1] (e11) {$e_1(y_1)$};
		\node[syn_stat, below=of e11] (e21) {$e_2(y_1)$};
		\node[below=0.2cm of e21] (vdots1) {$\vdots$};
		\node[syn_stat, below=0.2cm of vdots1] (eL1) {$e_L(y_1)$};
		\node[agg_stat_pass, below=of eL1] (ebar1) {$\bar{e}(y_1)$};
		\draw[arrow] (eL1.south) -- (ebar1.north);
		
		\node[syn_stat, below=of y2] (e12) {$e_1(y_2)$};
	\node at ($(ydots |- e12)$) (e1dots) {$\cdots$};
		\node[syn_stat, below=of e12] (e22) {$e_2(y_2)$};
	\node at ($(ydots |- e22)$) (e2dots) {$\cdots$};
		\node[below=0.2cm of e22] (vdots2) {$\vdots$};
		\node[syn_stat, below=0.2cm of vdots2] (eL2) {$e_L(y_2)$};
	\node at ($(ydots |- eL2)$) (eLdots) {$\cdots$};
		\node[agg_stat_fail, below=of eL2] (ebar2) {$\bar{e}(y_2)$};
	\node at ($(ydots |- ebar2)$) (ebardots) {$\cdots$};
		\draw[arrow] (eL2.south) -- (ebar2.north);
		
		\node[syn_stat, below=of ym] (e1m) {$e_1(y_m)$};
		\node[syn_stat, below=of e1m] (e2m) {$e_2(y_m)$};
		\node[below=0.2cm of e2m] (vdotsm) {$\vdots$};
		\node[syn_stat, below=0.2cm of vdotsm] (eLm) {$e_L(y_m)$};
		\node[agg_stat_pass, below=of eLm] (ebarm) {$\bar{e}(y_m)$};
		\draw[arrow] (eLm.south) -- (ebarm.north);
		
		
		
	\draw[brace,decorate] ([yshift=0.7cm]y1.north west) -- ([yshift=0.7cm]ym.north east) node[midway,above,yshift=10pt] {Candidate Space $\mathcal{Y}$};
	\draw[brace, decorate] ([xshift=-0.9cm]eL1.south west) -- ([xshift=-0.9cm]e11.north west) node[midway,left,xshift=-10pt] {Synthetics};
	\node at ([xshift=-5cm] ebar2.west) {Aggregation};
	\end{tikzpicture}
	\caption{A schematic illustration of the SAT procedure using synthetic e-values. For each candidate $y_i$, synthetic e-values are generated from the initial sets and are then aggregated. The final merged set, $\bar{\cC}_{\alpha}$, is constructed by including only those candidates whose aggregated statistic passes a pre-defined significance threshold. In this illustration, the aggregated statistics for $y_1$ and $y_m$ pass the threshold (shaded green), while the one for $y_2$ does not (shaded red).}
	\label{fig:sat_schematic_candidate}
\end{figure}

Our work makes several key contributions. First, we propose a principled framework that leverages the duality between hypothesis testing and set estimation to convert the set merging problem into one of aggregating test statistics, an approach that is robust to arbitrary dependencies among the input sets. Central to our framework is the concept of ``synthetic statistics", which mimic unknown oracle statistics while using only the initial sets and their coverage levels, thus bypassing any need for raw data. 
Second, the resulting merged set has a finite-sample theoretical coverage guarantee that is valid without any modeling assumptions.
Finally, we establish SAT's admissibility in the deterministic setting. This theoretical result provides foundational support for the commonly used voting heuristics. We establish not only that majority voting is a special case of our framework, but also that, more generally, any admissible procedure is equivalent to a weighted voting scheme. Our analysis therefore provides the first formal optimality guarantee for these popular methods, grounding a widely used family of heuristics in statistical admissibility theory.

Throughout the paper, we let $[n]=\{1,\dots,n\}$ for an integer $n\in\mathbb{Z}^+$, and let $\mathbb{R}^{\ge 0}$ be the set of non-negative real numbers. The indicator function is denoted by $\ind(\cdot)$, and $\mathbf{1}_L$ is an $L$-dimensional vector of ones. For two positive sequences $\{a_n\}$ and $\{b_n\}$, we write $a_n=O(b_n)$ if the ratio $a_n / b_n$ is bounded. The $L_1$ and $L_2$ norms of a function $f$ over the unit interval are denoted by $\|f\|_{L_1([0,1])}=\int_0^1|f(x)|\rd x$ and $\|f\|_{L_2([0,1])}=\{\int_0^1f(x)^2\rd x \}^{1/2}$, respectively.

\section{Deterministic Set Merging with Synthetic e-values}\label{sec:synthetics_SAT}
\subsection{Synthetic e-value}
	We start with deterministic set merging procedures using synthetic e-values in this section, and will move to randomized processes with synthetic p-values in the following section.

	Our procedure is motivated by the duality between set estimation and hypothesis testing, where a confidence set is formed by the collection of parameter values for which a null hypothesis is not rejected \citep{casella2024statistical}. This principle allows us to reframe the problem of merging sets as one of aggregating test statistics.
	
	Recently, the e-value has gained popularity as a measure of evidence for hypothesis testing \citep{vovk2021values, shafer2021testing}. An e-value is a non-negative random variable whose expectation is at most one under the null hypothesis. A key property is that for any $\alpha \in (0, 1)$, the test that rejects the null hypothesis if and only if $e \ge 1/\alpha$ controls the Type I error at level $\alpha$.	
	This testing rule provides a natural way to construct a synthetic e-value from an existing uncertainty set. While the true ``oracle e-value" used to form the initial set is unavailable in our data-light setting, we have the pair $(\cC_{\ell},\alpha_\ell)$ itself. 
	The set $\cC_{\ell}$ is, by construction, the non-rejection region of its underlying test. To align with the e-value framework, which rejects a hypothesis when $e \ge 1/\alpha_\ell$, we construct a statistic that emulates this behavior. Specifically, we assign a value of $1/\alpha_\ell$ to any point $y$ in the rejection region, i.e., $y \notin \cC_{\ell}$, and zero otherwise. This directly motivates defining the synthetic e-value as:


	\begin{equation}
		e_\ell(y) = \alpha_\ell^{-1}\cdot\ind(y\notin \cC_{\ell}),\quad\forall y\in\cY.
		\label{eq:sye_pred}
	\end{equation}
 The function $e_\ell(\cdot)$ is referred to as an e-function. The validity of this construction is established in the following proposition.
		\begin{proposition}
		{Suppose \eqref{eq:miscoverage} holds.
			Then, we have $\EE\{e_\ell(Y)\} \leq 1$ for all $\ell\in[L]$, where $e_\ell(\cdot)$ is the e-function defined in \eqref{eq:sye_pred}.}
		\label{thm:sye_pred}
	\end{proposition}

\subsection{Aggregating Synthetic e-values and Test Inversion}\label{sec:merge_e}
After constructing synthetic e-values for each initial set, the next step is to aggregate them into a single statistic, $\bar{e}(y)$. The appropriate aggregation method depends on the dependence structure of the initial sets.

We first consider the case where the synthetic e-values are mutually independent. This assumption holds, for instance, when $Y$ is a fixed parameter and each study independently collects data. A general aggregator for a vector $\mathbf{e} = (e_1,\ldots, e_L)$ of independent e-values is the average of products \citep{vovk2021values},
\begin{equation}
	\bar{e}_k = \binom{L}{k}^{-1}\sum_{\cI_k\in \mathcal{B}_k} \prod_{\ell\in \cI_k} e_\ell,
	\label{eq:combine_ind_e}
\end{equation}
where $\mathcal{B}_k$ is the set of all $k$-element subsets of $[L]$. When $k$ is fixed, we abbreviate $\bar{e}_k$ to $\bar{e}$. As shown in the following proposition, if the initial e-values are valid, so is their aggregation.

    \begin{proposition}\label{thm:agg_e_ind}
		Suppose $\EE(e_\ell(Y))\leq 1$ for all $\ell\in [L]$. Let $\bar{e}_k(Y) := G_e\{\eb(Y);k\}$ be defined as in \eqref{eq:combine_ind_e}.
		If the entries of $\eb(Y)$ are mutually independent, then $\EE\{\bar{e}_k(Y)\} \leq 1$ for any pre-determined $k\in[L]$.
\end{proposition}

If $Y$ is random, however, the synthetic e-values $e_\ell(Y)$ are generally not independent, as they share a common source of randomness. For this case of arbitrary dependence, a valid aggregation method is a convex combination.

  \begin{proposition}\label{thm:agg_e_dep}
  	Suppose $\EE(e_\ell(Y))\leq 1$ for all $\ell\in [L]$. Let $\bar{e}(Y):= G_e\{\eb(Y);\lambda\} =\sum_{\ell=1}^{L}\lambda_\ell e_\ell(Y)$, where $\lambda_\ell\geq 0$ for all $\ell$ and $\sum_{\ell=1}^{L}\lambda_\ell=1$.
	Then, we have $\EE\{\bar{e}(Y)\} \leq 1$.
\end{proposition}

The final step is to transform the aggregated e-value $\bar{e}(y)$ back into a merged uncertainty set. This is achieved through test inversion \citep{casella2024statistical}. An e-value $e(Y)$ controls the Type I error at level $\alpha$ using the test that rejects when $e(Y) \ge 1/\alpha$. Inverting this rule yields a $(1-\alpha)$-level uncertainty set. The following proposition formalizes this for our setting.

	\begin{proposition}\label{thm:alg_pred}
 For any $\alpha\in(0,1)$, define $\cC_\alpha = \{y\in\cY: e(y)<\tau/\alpha\}$. If $\EE\{e(Y)\} \leq 1$, then $\cC_\alpha$ is a  $(1-\alpha/\tau)$-level uncertainty set for $Y$.
\end{proposition}

By combining Propositions \ref{thm:sye_pred} - \ref{thm:alg_pred}, we see that for any pre-specified level $\alpha$, the final merged set can be constructed as
$$ \bar{\cC}_\alpha = \{ y \in \cY : \bar{e}(y) < \tau / \alpha \},$$
where $\bar{e}(y)$ is a valid aggregated e-value.
The complete Synthetics, Aggregation, and Test inversion (SAT) procedure is summarized in Algorithm \ref{alg:pred}. It provides a unified framework applicable to both the deterministic approach using e-values, as discussed, and a randomized version using synthetic p-values, which we introduce in Section \ref{sec:synthetics_SAT_p}.
The validity of SAT is formalized in the following theorem.

\begin{algorithm}[h] 
	\caption{The SAT Procedure} 
	\label{alg:pred}
	\begin{algorithmic}[1]
		\renewcommand{\algorithmicrequire}{\textbf{Input:}}
		\REQUIRE The pairs $\{(\cC_{\ell},\alpha_\ell)\}_{\ell\in[L]}$, candidate space $\cY$, a suitable aggregation function $G_e(\cdot)$ (or $G_p(\cdot)$), a control level $\alpha\in(0,1)$, an adjustment factor $\tau\in(0,1]$ (optional,  $\tau$=1 by default).
		\renewcommand{\algorithmicrequire}{\textbf{Initialize:}}
		\REQUIRE $\bar{\cC}_{\alpha}\leftarrow\{\}$.
		\FOR{each candidate $y\in\cY$}
		\FOR{each study $\ell\in[L]$}
		\STATE Generate synthetic e-values $e_\ell(y)$ by \eqref{eq:sye_pred}			
		(or synthetic p-values $p_\ell(y)$ by \eqref{eq:syp_pred}).
		
		\ENDFOR
		\STATE Calculate $\bar{e}(y) = G_e\{\eb(y)\}$ where $\eb(y)=\{e_1(y),\ldots,e_L(y)\}$
		\newline\\[-10pt]\hskip1.8cm(or $\bar{p}(y) = G_p\{\pb(y)\}$ where $\pb(y)=\{p_1(y),\ldots,p_L(y)\}$).
		\STATE Update $\bar{\cC}_{\alpha}\leftarrow\bar{\cC}_{\alpha}\cup\{y\}$, if $\bar{e}(y) < \tau/\alpha$ (or if $\bar{p}(y) > \alpha$).
		\ENDFOR
		\renewcommand{\algorithmicrequire}{\textbf{Output:}}
		\REQUIRE Merged set $\bar{\cC}_{\alpha}$.	
	\end{algorithmic} 
\end{algorithm}

\begin{theorem}\label{thm:main_validity}
	Suppose the initial uncertainty sets $\{\cC_\ell\}_{\ell=1}^L$ satisfy \eqref{eq:miscoverage}. Let the synthetic statistics be generated by \eqref{eq:sye_pred} or \eqref{eq:syp_pred}, and let the aggregation function be one that produces a valid aggregated e-value or p-value, as established in Propositions \ref{thm:agg_e_ind}--\ref{thm:agg_e_dep} and \ref{thm:agg_p_dep}--\ref{thm:agg_p_ind}. Then the merged set $\bar{\cC}_\alpha$ produced by Algorithm \ref{alg:pred}, with the default choice of $\tau=1$, is a valid $(1-\alpha)$-level uncertainty set satisfying $\PP(Y \in \bar{\cC}_\alpha) \ge 1-\alpha$ for any $\alpha \in (0,1)$.
	\end{theorem}

\begin{remark}\label{rmk3}
	The SAT procedure with synthetic e-values and a convex combination is equivalent to a weighted voting scheme. A point $y$ is included in the merged set if $\sum_{\ell=1}^L \lambda_\ell e_\ell(y) < \tau/\alpha$. Substituting the definition of synthetic e-values, this condition becomes
	$
	\sum_{\ell=1}^L \frac{\lambda_\ell}{\alpha_\ell} \{\ind(y \notin \cC_\ell)\} < \frac{\tau}{\alpha}.
	$
	Here, each set $\cC_\ell$ casts a ``vote" for including $y$, and the vote is weighted by $\lambda_\ell/\alpha_\ell$.
	When $\alpha_\ell = \alpha/2$ for all $l \in [L]$ and an arithmetic mean is used for aggregation ($\lambda_\ell = 1/L$), the majority voting procedurse of \cite{gasparin2024merging} appears as a special case.
\end{remark}

\subsection{Admissibility of SAT for Deterministic Set Merging}\label{sec:ad}
In the previous subsection, we formally introduced the SAT procedure for deterministic set merging.
However, it remains an open question whether the SAT procedure can be strictly improved. This naturally leads to the concept of admissibility, which provides a formal criterion for optimality. We first give the definition of admissibility below.

\begin{definition}\label{func:setmerge}
	A function $f$ is called a level-$\alpha$ deterministic set merging function if it takes a collection of $L$ uncertainty sets along with their miscoverage guarantee $\{(\cC_{1},\alpha_1),\ldots, (\cC_{L},\alpha_L)\}$ and outputs a single set $\cC_\alpha$ that satisfies $\mathbb{P}(Y\notin \cC_\alpha)\leq \alpha$, without introducing any external randomness. A level-$\alpha$ deterministic set merging function $f$ is called \textit{admissible} if there does not exist another level-$\alpha$ deterministic set merging function $g\neq f$ such that $g(\cdot)\subseteq f(\cdot)$ for any valid inputs.
	
\end{definition} 

The following theorem establishes that the SAT procedure is not just a valid method, but that its structure is necessary for any admissible approach under general dependence.

\begin{theorem}\label{admissibility}
Under general dependence, every admissible level-$\alpha$ deterministic set merging function can be represented in the form of SAT (Algorithm \ref{alg:pred}), with synthetic e-value and a convex combination in the aggregation step.
\end{theorem}

\begin{remark}
	The converse of Theorem \ref{admissibility} is not true. In Remark \ref{counterexample} of the Supplement we give a counterexample where SAT with synthetic e-values and certain choice of convex combination is not admissible.
\end{remark}

The main idea in proving Theorem \ref{admissibility} is to exploit the duality between uncertainty sets and the e-function defined in \eqref{eq:sye_pred}. 
This duality allows us to recast the problem of merging uncertainty sets as that of merging e-functions. This perspective enables us to draw on established results from the e-value literature for deeper insight. 
A key step in the proof of Theorem \ref{admissibility} is to show that all admissible e-function mergers that {map} collections of e-functions to a single e-function must take the form of convex combinations. We emphasize that this result is not a direct consequence of Theorem 1 in \cite{wang2025only}, as our focus here is on merging functions of a specific form rather than general random variables. In fact, our proof of Theorem \ref{admissibility} uses different techniques from those in the existing e-value literature.

In the special case where all $L$ initial uncertainty sets share the same miscoverage level, it is natural to consider symmetric set merging functions—those that are invariant under permutations of the input sets. The following theorem establishes the admissibility of SAT under this setting. 
It implies that the merged set produced by any valid and symmetric deterministic function must contain the set generated by the SAT procedure with an arithmetic mean aggregator.
\begin{theorem}\label{admissibility2}
	Under the assumption that all $L$ initial uncertainty sets have the same miscoverage level,  the SAT procedure with synthetic e-values and arithmetic mean aggregation yields the only admissible symmetric deterministic set merging function.
\end{theorem}

When the initial sets are known to be independent and $Y$ is a fixed parameter, the characterization of admissible procedures becomes more intricate. Unlike the dependent case where a single class of functions (convex combinations) is admissible, here the optimal function's structure depends on the specific values of $(\alpha_1, \ldots, \alpha_L)$ and $\alpha$.
	
The core of the challenge is the discrete nature of the problem. When constructing an admissible function, a key step is to find a function $F: \prod_{\ell=1}^{L} \{0, 1/\alpha_\ell\} \to \{0, 1/\alpha\}$ that ``exhausts the error budget", i.e., $\EE\{F(\pmb{e}^\dagger)\} = 1$, 
	where $\pmb{e}^\dagger$ is the random vector with independent components $e^\dagger_\ell$ taking value $1/\alpha_\ell$ with probability $\alpha_\ell$ and $0$ otherwise.
	The output of $F$ can be written as $F(\pmb{e}) = \alpha^{-1} \ind\{\pmb{e} \in A\}$ for some set $A\subseteq \prod_{\ell=1}^{L} \{0, 1/\alpha_\ell\}$. The condition $\EE\{F(\pmb{e}^\dagger)\} = 1$ is thus equivalent to $\PP(\pmb{e}^\dagger \in A) = \alpha$. However, the set of attainable probabilities for this event is a finite, discrete set determined by sums of products of the $\alpha_\ell$. If the target level $\alpha$ does not belong to this set, no such function $F$ exists, and a simple sufficient condition for admissibility cannot be applied.

Despite this complexity, a complete characterization of admissibility for the independent case is possible. An admissible merging function corresponds to an optimal rule for partitioning the $2^L$ possible outcomes of the synthetic e-values. A merging function is admissible if the set $A$ cannot be enlarged any further without violating the procedure's validity.
 This leads to a clear condition: a function is inadmissible if there is an outcome currently not in $A$ but can be moved to $A$ while the function remains valid. 
	
While this provides a theoretical criterion, it confirms that the optimal function depends on the entire discrete probability distribution defined by the $\alpha_\ell$ and $\alpha$ values. The full technical details and a formal statement are provided in Section \ref{ap:merge_indep_no_random} of the Supplement. This nuanced theoretical landscape explains why there is no single, uniformly optimal aggregator for the independent case. Despite this ambiguity, our simulation studies suggest that the SAT procedure with the average of pairwise products ($k=2$ in \eqref{eq:combine_ind_e}) provides an effective and powerful approach in practice.

	\section{Randomized Set Merging with Synthetic p-values}\label{sec:synthetics_SAT_p}
	\subsection{Synthetic p-values}
	
	While the deterministic approach using e-values provides strong admissibility guarantees, randomized procedures can offer advantages in certain settings, often yielding smaller merged sets. A more traditional test statistic that naturally accommodates randomization is the p-value. 	A p-value $ p \in [0,1] $ is a random variable that satisfies $ \PP(p \leq t) \leq t $ for all $ t \in (0,1) $ under the null hypothesis.
As with e-values, p-values can be inverted to form uncertainty sets. In our data-light setting, however, the true ``oracle p-value" used to generate the initial set is unavailable. We therefore construct a randomized synthetic p-value that mimics the behaviour of the oracle. For each uncertainty set $\cC_{\ell}$ with control level $\alpha_\ell$, we define the synthetic p-value as
	\begin{equation}
		\label{eq:syp_pred}
		p_\ell(y):=p_\ell(y;\cC_{\ell},\alpha_\ell)\sim\Unif\left(0,\alpha_\ell\right)\cdot\ind(y\notin \cC_{\ell})+\Unif\left(\alpha_\ell,1\right)\cdot\ind(y\in \cC_{\ell}),\quad\forall y\in\cY.
	\end{equation}
The construction is intuitive: if a point $y$ is outside the set $\cC_\ell$, the oracle p-value was likely small, so we draw a value from $\Unif(0, \alpha_\ell)$. Conversely, if $y$ is inside the set, the oracle p-value was likely large, and we draw from $\Unif(\alpha_\ell, 1)$. The following proposition confirms that this construction yields a valid, super-uniform p-value.
	\begin{proposition}
		Suppose \eqref{eq:miscoverage} holds.
		Then, we have $\PP\{p_\ell(Y)\leq t\}=t+\Delta_\ell(t)$ for all $t\in[0,1]$ and $\ell\in[L]$, where
		$p_\ell(\cdot)$ is the mapping defined in \eqref{eq:syp_pred} and
		$
		\Delta_\ell(t) = \left\{1-\PP\left(Y\notin\cC_{\ell}\right)/{\alpha_\ell}\right\}\cdot\{(t-\alpha_\ell)\ind\left(t>\alpha_\ell\right)-(t-t\alpha_\ell)\}/(1-\alpha_\ell)\leq0.
		$
		\label{thm:syp_pred}
	\end{proposition}
	
	\begin{remark}\label{rmk1}
		As stated in the above proposition, exact uniformity ($\PP\{p_\ell(Y)\leq t\}=t$) is achieved if the uncertainty set $\cC_{\ell}$ is exact, i.e., $\PP(Y\notin \cC_{\ell})=\alpha_\ell$. 
	\end{remark}
	
	\begin{remark}\label{rmk2}
		Given an uncertainty set derived from data, the synthetic p-values defined in \eqref{eq:syp_pred} are randomly generated. The sensitivity of the merged set to this randomness is investigated in Section \ref{ap:sensitivity} of the Supplement. 
	\end{remark}

\subsection{Aggregating Synthetic p-values and Test Inversion}\label{sec:merge_p}
Having constructed synthetic p-values, the next step is to combine them into a single aggregated p-value, $\bar{p}(y)$. The choice of aggregation method depends on the dependence structure of the p-values.

We first address the general case of arbitrary dependence. 
Let $\mathbf{p} = (p_1,\ldots,p_L)$ be a vector of synthetic p-values.
If each $p_\ell$ is a conventional p-value, then as pointed out in \cite{vovk2022admissible}, we can define an aggregation function $G_p:[0,1]^L\mapsto[0,1]$ through the corresponding rejection regions. 
More precisely, given any increasing collection of Borel lower sets $\{R_\alpha \subseteq [0,1]^L:\alpha\in (0,1)\}$, if $\PP(\mathbf{p}\in R_\alpha)\leq \alpha$ for any $\alpha\in (0,1)$ under the null hypothesis, then
$G_p(\mathbf{p}) = \inf \{ \alpha \in (0, 1) : \mathbf{p} \in R_\alpha \}$
defines a valid p-value aggregation function. 
Synthetic p-values can be aggregated using the same idea.

Consider a family of regions defined as follows:
\begin{equation}
	R_\alpha=\left\{\mathbf{p} \in[0,1]^L: \sum_{\ell=1}^{L} \lambda_\ell\cdot f_\ell\left(\frac{p_\ell}{\alpha}\right) \geq 1\right\},
	\label{eq:combine_dep_general}
\end{equation}
where $\lambda_\ell$'s are non-negative numbers satisfying $\sum_{\ell=1}^{L}\lambda_{\ell}=1$, and $f_\ell$'s are p-to-e calibrators. {Here, p-to-e calibrator is a decreasing function $f: [0, \infty) \mapsto [0, \infty]$ such that $\|f\|_{L_1([0,1])}\leq1$ \citep{vovk2021values,gasparin2024combining}.}
This form of p-value aggregation is proposed in \cite{vovk2022admissible}.
If $p_\ell$'s are conventional p-values, $G_p(\mathbf{p}) = \inf \{ \alpha \in (0, 1) : \mathbf{p} \in R_\alpha \}$ with $R_\alpha$ defined in \eqref{eq:combine_dep_general}  encompasses some popular aggregation methods.
For instance, if $f_\ell(p) = 2 - 2p $ and $\lambda_\ell=1/L$ for all $\ell\in[L]$, we get the arithmetic mean aggregation function $G_p(\mathbf{p}) = 2\sum_{\ell=1}^Lp_\ell/L$; for a pre-determined $k\in[L]$, if  $f_\ell(p)=L/k\cdot\ind\{p\in(0,k/L)\}+\infty \ind(p=0)$  and $\lambda_\ell=1/L$ for all $\ell\in[L]$, we obtain the R\"uger's method $G_p(\mathbf{p})=L/k\cdot p_{(k)}$.  The following proposition establishes the super-uniformity of the aggregated synthetic p-values.

\begin{proposition}\label{thm:agg_p_dep}
	Suppose $\PP\{p_\ell(Y)\leq t\}\leq t$ for all $\ell\in[L]$. Let $\bar{p}(Y):=G_p\{\pb(Y)\} = \inf \{ \alpha \in (0, 1) : \mathbf{p}(Y) \in R_\alpha \}$, where $R_\alpha$ is defined in \eqref{eq:combine_dep_general}.
		Then, we have $\PP\{\bar{p}(Y) \leq t\} \leq t$ for all $t\in [0,1]$.
\end{proposition}

When the synthetic p-values are known to be independent, more powerful aggregation methods are available. Specifically, we consider the rejection regions of the following form:

\begin{equation}
	R_\alpha=\left\{\pb \in [0,1]^L: \sum_{\ell=1}^L S_\ell(p_\ell) \geq  c_{1-\alpha}(\{S_\ell\}_{\ell\in[L]})\right\},
	\label{eq:combine_ind_general}
\end{equation}
where $S_\ell:[0,1]\mapsto\RR$ is decreasing and $c_{1-\alpha}(\{S_\ell\}_{\ell\in[L]}) = \mathrm{Quantile}(1-\alpha; \sum_{\ell=1}^L S_\ell(U_\ell))$ with $U_\ell \iid \Unif(0,1)$. 
Notably, \eqref{eq:combine_ind_general} encompasses some of the most widely used aggregation methods for the conventional p-values. 
For example, by setting $S_\ell(t)=-2\log t$, we obtain Fisher's aggregation function $G_p(\pb)=1-F_{\chi^2_{2L}}(-2\sum_{\ell=1}^L\log p_\ell)$, where $F_{\chi^2_{2L}}$ denotes the CDF of a centered $\chi^2$-random variable with $2L$ degrees of freedom \citep{fisher1948};
by taking $S_\ell(t)=-\lambda_\ell \cdot \Phi^{-1}(t)$, where $\lambda_\ell$'s are some positive constants, we obtain the Lipt{\'a}k's method $G_p(\pb)=\Phi\Big\{\sum_{\ell=1}^L \lambda_\ell \cdot \Phi^{-1}(p_\ell) / \sqrt{\sum_{\ell=1}^L \lambda_\ell^2}\Big\}$ \citep{liptak1958combination}. 
For detailed comparisons of various aggregation methods under independence, see, for example, \cite{heard2018choosing}.

The following proposition shows that the aggregated synthetic p-value via \eqref{eq:combine_ind_general} is still marginally super-uniform.

\begin{proposition}\label{thm:agg_p_ind}
Suppose $\PP\{p_\ell(Y)\leq t\}\leq t$ for all $\ell\in[L]$. Let $\bar{p}(Y):=G_p\{\pb(Y)\} = \inf \{ \alpha \in (0, 1) : \mathbf{p}(Y) \in R_\alpha \}$, where $R_\alpha$ is defined in \eqref{eq:combine_ind_general}.
		If the entries of $\pb(Y)$ are mutually independent, then $\PP\{\bar{p}(Y) \leq t\} \leq t$ for all $t\in [0,1]$.
\end{proposition}

The final step of the randomized procedure is to convert the aggregated p-value, $\bar{p}(y)$, back into a merged set via test inversion. A p-value controls the Type I error by rejecting the null hypothesis when $p \le \alpha$. Inverting this decision rule directly yields the corresponding uncertainty set.


\begin{proposition}\label{thm:alg_pred_p}
For any $\alpha\in(0,1)$, define $\cC_\alpha = \{y\in\cY: p(y)>\alpha\}$. If $\PP\{p(Y)\leq t\} \leq t$ for all $t\in[0,1]$, then $\cC_\alpha$ is a $(1-\alpha)$-level uncertainty set for $Y$.
		\end{proposition}
	Applying Proposition \ref{thm:alg_pred_p}, the final merged set is constructed as
$$ \bar{\cC}_\alpha = \{ y \in \cY : \bar{p}(y) > \alpha \},$$
where $\bar{p}(y)$ is a valid aggregated p-value from either \eqref{eq:combine_dep_general} or \eqref{eq:combine_ind_general}. 

The validity of the above $\bar{\cC}_\alpha$ is a direct consequence of the validity of its constituent steps, and is formalized in Theorem \ref{thm:main_validity} in the previous section.


  \subsection{SAT Procedure for Infinite Candidate Space}

In many cases the candidate space $\cY$ is infinite or even uncountable, which makes it impractical to compute $\{p_\ell(y)\}_{y\in\cY}$ individually as specified in Algorithm \ref{alg:pred}. 
To address this, we propose a modified version of SAT in Algorithm \ref{alg:pred_imp}. 
Specifically, we split $\cY$ into non-overlapping subsets and select a representative candidate from each subset to execute Algorithm \ref{alg:pred}.  
Note that Algorithm \ref{alg:pred_imp} is guaranteed to terminate in finite time if $L<\infty$ and $\cC_{\ell}$ are finite unions of connected sets. 
The validity of Algorithm \ref{alg:pred_imp} is given by the next theorem.
\begin{theorem}\label{thm:practice}
	Algorithm  \ref{alg:pred_imp} has the same theoretical guarantee as Algorithm \ref{alg:pred}.
\end{theorem}
A proof of Theorem \ref{thm:practice} is given in Section \ref{ap:thm_practice} of the Supplement.

\begin{algorithm}[h] 
	\caption{The SAT Procedure for Practical Implementation} 
	\label{alg:pred_imp}
	\begin{algorithmic}[1]
		\renewcommand{\algorithmicrequire}{\textbf{Input:}}
		\REQUIRE The pairs $\{(\cC_{\ell},\alpha_\ell)\}_{\ell\in[L]}$, candidate space $\cY$, a suitable aggregation function $G_e(\cdot)$ (or $G_p(\cdot)$), a control level $\alpha\in(0,1)$, an adjustment factor $\tau\in(0,1]$ (optional, take $\tau$=1 as default).
		\renewcommand{\algorithmicrequire}{\textbf{Initialize:}}
		\REQUIRE $\bar{\cC}_{\alpha}\leftarrow\{\}$, $\mathcal{M}_0\leftarrow\{\cY\}$.	
		\FOR{each $\ell$ in [L]}
		\STATE Iteratively split $\cY$ into $\mathcal{M}_\ell = \cap\{\mathcal{M}_{\ell-1}, \cC_{\ell}\} \cup \setminus\{\mathcal{M}_{\ell-1}, \cC_{\ell}\}$ for all $\ell\in[L]$, where $\cap\{\cA,b\} = \{a\cap b: a \in \cA\}$ and $\setminus\{\cA,b\} = \{a\setminus b: a \in \cA\}$.
		\ENDFOR	
		
		\FOR{each $\tilde{\cY} \in \mathcal{M}_L$}\label{line:division}
		\STATE Select any representative candidate $y\in\tilde{\cY}$.
		\FOR{each study $\ell\in[L]$}
		\STATE Generate synthetic e-values $e_\ell(y)$ using \eqref{eq:sye_pred} (or synthetic p-values $p_\ell(y)$ using \eqref{eq:syp_pred}).
		\ENDFOR
		\STATE Calculate $\bar{e}(y) = G_e\{\eb(y)\}$ where $\eb(y)=\{e_1(y),\ldots,e_L(y)\}$.
		\newline\\[-10pt]\hskip1.8cm(or $\bar{p}(y) = G_p\{\pb(y)\}$ where $\pb(y)=\{p_1(y),\ldots,p_L(y)\}$).
		\STATE Update $\bar{\cC}_{\alpha}\leftarrow\bar{\cC}_{\alpha}\cup\tilde{\cY}$, if $\bar{e}(y) < \tau/\alpha$ (or if $\bar{p}(y) > \alpha$).
		\ENDFOR
		\renewcommand{\algorithmicrequire}{\textbf{Output:}}
		\REQUIRE Merged set $\bar{\cC}_{\alpha}$.	
	\end{algorithmic} 
\end{algorithm}

	\section{Content of the Merged Set}

	While admissibility provides a clear criterion for optimality in deterministic merging, a similar formal framework for randomized procedures is less straightforward. We therefore turn our attention to a different measure of performance: the content of the merged set. Specifically, we analyze the probability that a candidate point $y$ is excluded from the final set, $\PP(y \notin \bar{\cC}_\alpha)$. This quantity is key to understanding the set's size and composition, as the expected size of $\bar{\cC}_\alpha$ is $\int_{\cY} \PP(y \in \bar{\cC}_\alpha) \, \mu(\rd y)$, where $\mu$ is a suitable measure.
	
In this section we focus on the case where the synthetic statistics are independent; an analysis for dependent scenarios is provided in Section  \ref{sec:size_dependence} of the Supplement. We begin with the randomized procedure based on synthetic p-values.
	 \begin{assumption}
		Assume that $S_\ell = S$ for all $\ell \in [L]$ in \eqref{eq:combine_ind_general} and $S : [0,1] \mapsto \mathbb{R}$ satisfies:
		\begin{itemize}
			\setlength{\itemsep}{-3pt}
			\item [(\romannumeral1)] $\|S\|_{L^2([0,1])} \leq C_S$ for some constant $C_S>0$.
			\item [(\romannumeral2)] $\alpha^{-1} \int_0^\alpha S(t) \, \rd t > (1-\alpha)^{-1} \int_\alpha^1 S(t) \, \rd t$ for all $\alpha \in (0,1)$.
		\end{itemize}
		\label{as:combine_ind}
	\end{assumption}
	
	We remark that Part $(\romannumeral2)$ of  Assumption \ref{as:combine_ind} is automatically satisfied if $S$ is strictly decreasing. Assumption \ref{as:combine_ind} is met by common choices of $S(t)$, such as $S(t)=-2\log (t)$ and $S(t)=-\Phi^{-1}(t)$. 
	We establish the results on $\PP(y \notin \bar{\cC}_\alpha)$ in the following theorem.
	\begin{theorem}
		Suppose \eqref{eq:miscoverage} holds with $\alpha_\ell=\alpha$ for all $\ell\in [L]$, and $\cC_{\ell,\alpha}$'s are independent and identically distributed. Let $\bar{p}(y)= \inf \{ \alpha \in (0, 1) : \pb(y) \in R_\alpha \}$ for each $y\in\cY$, where $R_\alpha$ is defined in \eqref{eq:combine_ind_general}.
		If $S_\ell$ satisfies Assumption \ref{as:combine_ind} and $\Var[S\{p_1(y)\}]>0$ for all $y\in\cY$, then for any $\alpha'\in (0,1)$, when $L\rightarrow\infty$, there exists constant $C>0$ such that
		$$
		\PP(y\notin\bar{\cC}_{\alpha'})=1-O\{\exp(-C \cdot L)\}, \quad\forall y\in \{y\in\cY:\PP(y\notin \cC_{1,\alpha})>\alpha\},
		$$
		where $\bar{\cC}_{\alpha'}=\{y\in\cY: \bar{p}(y)>\alpha'\}$.
		\label{thm:ind_power}
	\end{theorem}
	
	To interpret Theorem \ref{thm:ind_power}, we consider the following simple example. Suppose $Y=\theta^*$ is a fixed unknown parameter, each of the $L$ studies independently draws a sample $X_\ell$ from $\cN(\theta^*,1)$ and constructs an uncertainty set $\cC_{\ell,\alpha}=\{\theta\in\mathbb{R}: |\theta-X_\ell|\leq z_{1-\alpha/2}\}$. In this case, the set $ \{\theta\in\mathbb{R}:\PP(\theta\notin \cC_{1,\alpha})>\alpha\}$ is simply $\mathbb{R}\backslash\{\theta^*\}$ and Theorem \ref{thm:ind_power} implies that for all $\theta\neq \theta^*$, the probability that $\bar{\cC}_{\alpha'}$ includes $\theta$ converges to 0 at a rate of $\exp(-C \cdot L)$ for any $\alpha'\in (0,1)$.
	Consequently,  the final merged set will converge to the singleton $\{\theta^*\}$ if $L\rightarrow \infty$ and becomes infinitesimal in size.
	
	We next present the parallel result for synthetic e-value aggregation.
	\begin{theorem}
		Suppose \eqref{eq:miscoverage} holds with $\alpha_\ell=\alpha$ for all $\ell\in [L]$, and $\cC_{\ell}$'s are independent and identically distributed. Let $\bar{e}_k(y) = G_e\{\eb(y);k\}$ be defined as in \eqref{eq:combine_ind_e} for each $y\in\cY$.
		Then, for any fixed constant $k\in[L]$ and any $\alpha'\in (0,1)$, when $L\to\infty$, there exists a constant $C>0$ such that
		$$
		\PP(y\notin\bar{\cC}_{\alpha'})=1-O\{\exp(-C \cdot L)\},\quad \forall y\in \left\{y\in\cY:\PP(y\notin \cC_{1,\alpha})>\alpha\left(\frac{\tau}{\alpha'}\right)^{1/k}\right\},
		$$
		where $\bar{\cC}_{\alpha'}=\{y\in\cY: \bar{e}_k(y)<\tau/\alpha'\}$ for any fixed $\tau\in(0,1]$.
		\label{thm:ind_power_ev}
	\end{theorem}
		Comparing the results in Theorems \ref{thm:ind_power} and \ref{thm:ind_power_ev} highlights a nuance between the two approaches. 
		For the default choice of $\tau=1$, the set $\{y\in\cY:\PP(y\notin \cC_{1,\alpha})>\alpha(\tau/\alpha')^{1/k}\}$ is increasing in $k$, and is always a subset of $\{y\in\cY:\PP(y\notin \cC_{1,\alpha})>\alpha\}$ in Theorem \ref{thm:ind_power} for any fixed $k$. 	This suggests that SAT with synthetic p-values is less conservative than its e-value counterpart, as will be affirmed in the following numerical sections.

		\section{Simulation Studies}\label{sec:simulation}
\subsection{Experimental Setup}

		In this section, we study the empirical performance of the proposed methods on simulated datasets.  
		Throughout the paper, we abbreviate the methods using the format \texttt{($\cdot$)+($\cdot$)}. 
		The first component denotes the type of synthetic statistic: \texttt{SyE} and \texttt{SyP} correspond to the synthetic e-value and p-value, respectively. The variant \texttt{SyP(naïve)} refers to a direct construction of the synthetic p-value via \(\alpha_\ell \cdot \ind(y \notin \mathcal{C}_{\ell,\alpha_\ell}) + \ind(y \in \mathcal{C}_{\ell,\alpha_\ell})\), and it serves as a deterministic baseline for comparison with \texttt{SyE} procedures. Additionally, \texttt{OrP} denotes the “oracle p-value,” which is used to generate the initial uncertainty sets.
		The second part refers to the aggregation methods, which we summarize in Table \ref{tab:comb_mtd}. 
		In the last step of inverting the synthetic e-values, we set $\tau = 1$.
		When all initial uncertainty sets are independent of each other and have the same coverage level, we include the procedure in Section 2.6 of \cite{gasparin2024merging} for comparison, which is denoted as \texttt{MV Binom}.
		
		\begin{table}[h!]
			\centering
			\footnotesize
			\renewcommand\arraystretch{1.1} 
			\begin{tabular}{lllc}
				\toprule 
				&\bf Method  &\bf Abbreviation   &\bf Aggregation function            \\ 
				\midrule		
				\multirow{2.3}{*}{E-value} & {Arithmetic Mean} & \texttt{AM} & $\sum_{\ell=1}^Le_\ell/L$ \\
				& {Equation \eqref{eq:combine_ind_e} with $k=2$} & \texttt{U2} & $\binom{L}{2}^{-1}\sum_{\cI_2\in \mathcal{B}_2} \prod_{\ell\in \cI_2} e_\ell(Y)$ \\	[4pt]
				\hline
				\multicolumn{3}{l}{}\\[-10pt]
				\multirow{3}{*}{P-value} & {Fisher's method} & \texttt{Fisher} & $1-F_{\chi^2,{2L}}\big(-2\sum_{\ell=1}^L \log p_\ell\big)$   \\
				&{{Arithmetic mean}} & \texttt{AM}  & $2\sum_{\ell=1}^L p_\ell / L$   \\
				&{R\"uger's method with $k=1$}   & \texttt{R\"uger}    & ${L} \cdot p_{(1)} \cdot\ind(p_{(1)}>0)$ \\
				\bottomrule
			\end{tabular}
			\caption{{A summary of aggregation methods.}}
			\label{tab:comb_mtd}
		\end{table}

		
		The following four scenarios are considered.
		\begin{enumerate}[leftmargin=80pt]
			\setlength{\itemsep}{-3pt}
			\item[\emph{Scenario 1.}] $L=5$, $\alpha_1=\cdots=\alpha_L=\alpha/2$ varies from $0.01$ to $0.1$;
			\item[\emph{Scenario 2.}] $L$ varies from $2$ to $9$, $\alpha_1=\cdots=\alpha_L=\alpha/2=0.05$;
			\item[\emph{Scenario 3.}] $L=5$, $(\alpha_1,\ldots,\alpha_L)=(0.01,\ldots,0.05)$, $\alpha$ varies from $0.01$ to $0.1$;
			\item[\emph{Scenario 4.}] $L=5$, $\alpha_1=\cdots=\alpha_L$ varies from $0.01$ to $0.1$, $\alpha = 0.1$.
		\end{enumerate}
		For all of the settings, we simulate $ L \in \mathbb{N} $ initial uncertainty sets $ \mathcal{C}_{\ell,\alpha_\ell} $ for $ \ell \in [L] $ with individual set coverage guarantee: $\PP(Y\notin \mathcal{C}_{\ell,\alpha_\ell})\leq \alpha_\ell$. Our goal is to construct a merged set $\bar{\cC}_\alpha$ that satisfies $\PP(Y\notin \bar{\cC}_\alpha)\leq \alpha$. All experiments are based on 5000 replications, and the average results are reported.

		\subsection{Merging Independent Uncertainty Sets}\label{sec:simu1}
\begin{figure}[ht]
			\centering
			\includegraphics[width=\linewidth]{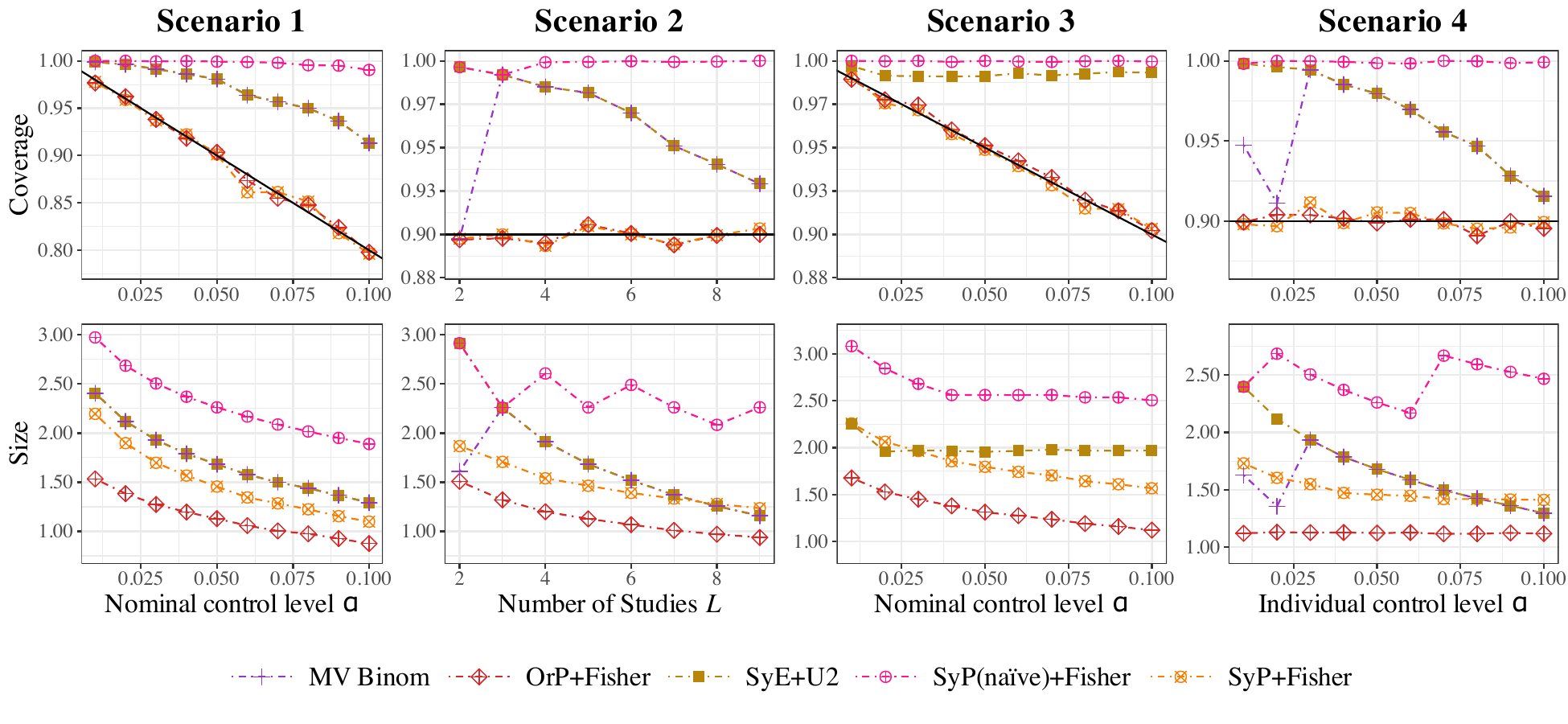}
			\caption[Normal-mean simulation]{\small Coverage and size of the merged uncertainty sets for the normal mean estimation problem. Each individual set is constructed based on a two-sided $z$-test.}
			\label{fig:sim_indep_normal}
		\end{figure}

		We let $Y=2$ be a fixed parameter. For each of the $L$ studies, we independently draw $n=3$ samples from $\cN(Y,1)$, and denote the mean of the $n$ samples obtained by study $\ell$ as $\bar{X}_{\ell}$. 
		The oracle p-value at each candidate point $y$ is computed as $p^{\text{or}}(y) = 2\Phi\left(-\sqrt{n} |y - \bar{X}_\ell|\right)$, and the corresponding uncertainty set $ \mathcal{C}_{\ell,\alpha_\ell} $ is constructed by
		$
		\mathcal{C}_{\ell,\alpha_\ell} = \left[\bar{X}_\ell - z_{\alpha_\ell/2}/\sqrt{n}, \bar{X}_\ell + z_{\alpha_\ell/2}/\sqrt{n}\right].$
		The comparison results of various methods are summarized in Figure \ref{fig:sim_indep_normal}. 
		
		From the top panel of Figure \ref{fig:sim_indep_normal}, we observe that all methods successfully control the desired coverage level. The size comparisons presented in the second row indicate that \texttt{SyP+Fisher} demonstrates clear advantages over other variations of SAT.
		Note that this does not contradict the theory presented in Section \ref{sec:ad}, as \texttt{SyP+Fisher} is a randomized merging procedure, whereas Theorems \ref{admissibility} and \ref{admissibility2} pertain exclusively to deterministic merging procedures. In contrast, \texttt{SyP(naïve)+Fisher} is deterministic. As shown in Figure \ref{fig:sim_indep_normal}, its resulting size is consistently larger than those of \texttt{SyE+U2}, which numerically confirms our theoretical results. 
		In addition, the observed size trends align with the theoretical results in Theorem \ref{thm:ind_power} and Theorem \ref{thm:ind_power_ev}.

		\subsection{Merging Dependent Uncertainty Sets}\label{sec:dependent_simu}
		Consider the linear model $Y = X^\top \beta + \epsilon$, where $ X \in \mathbb{R}^{150} $ is the covariate vector, $ \beta \in \mathbb{R}^{150} $ is an unknown vector of coefficients, and $ \epsilon $ is the error term that follows $ \mathcal{N}(0, 1) $. 
		In this experiment, we generate the first 10 entries of $ \beta $ from $ \mathcal{N}(0, 4I_{10 \times 10}) $ and set the remaining entries to 0. We then generate 400 pairs of data according to the linear model, with $ X $ sampled independently from $ \mathcal{N}(0, I_{150 \times 150}) $. Next, we sample one more $ X $ from $ \mathcal{N}(0, I_{150 \times 150}) $, and our goal is to produce an uncertainty set for the corresponding $ Y$.
		In this setting $Y$ is random, so the synthetic p-values and e-values are not independent. Consequently, aggregation methods like \texttt{U2} and \texttt{Fisher} are no longer valid, and we do not include them for comparison.

\begin{figure}[ht]
			\centering
			\includegraphics[width=\linewidth]{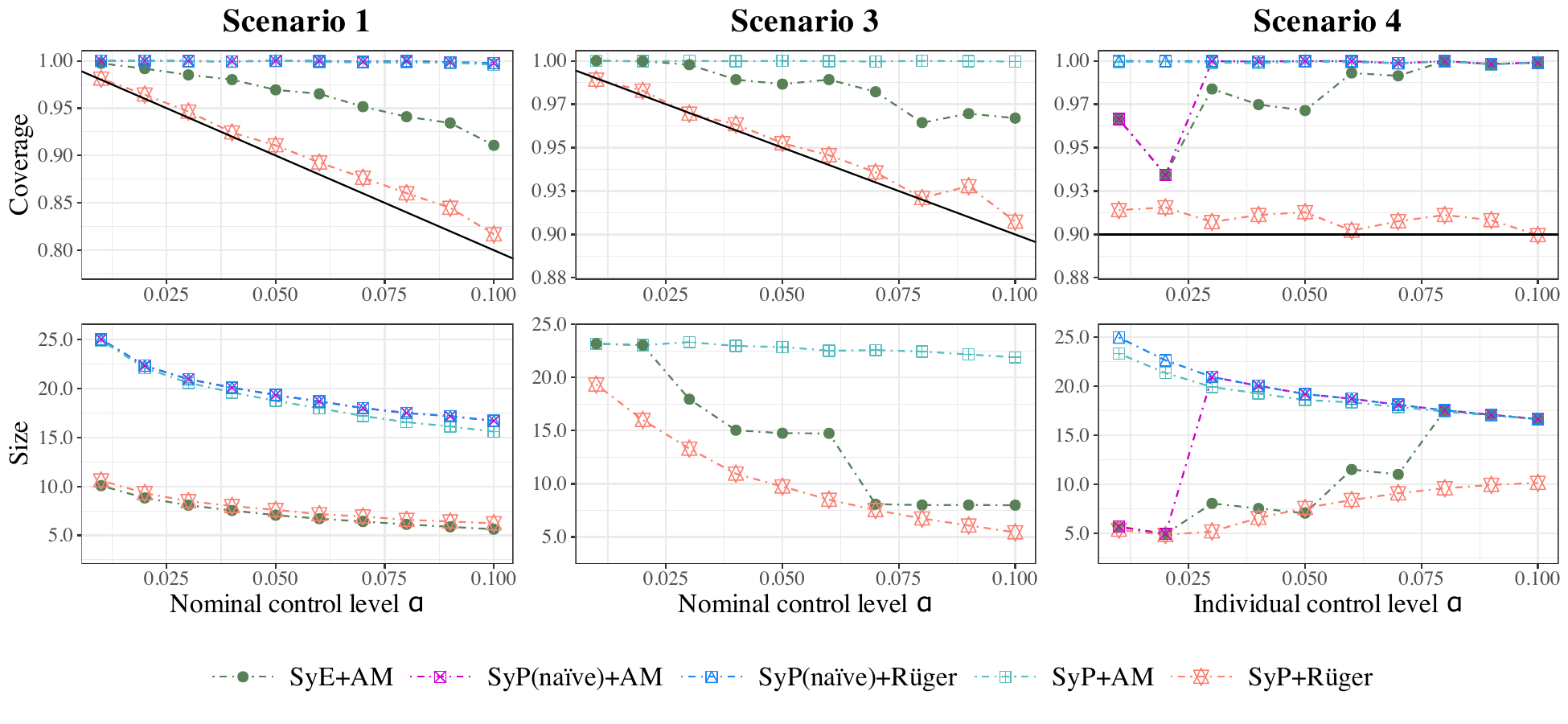}
			\caption[Full-conformal methods]{\small Coverage and size of the merged conformal prediction sets evaluated using different score functions and merging methods. The initial sets are constructed using a full conformal approach, with neural network, random forest, LASSO, and linear model selected as the score functions.}
			\label{fig:sim_dep_mmthd}
		\end{figure}

		\noindent\textbf{Conformal prediction set with different learning algorithms.}
		We consider the case where each study chooses different learning algorithms to construct conformal prediction sets. More precisely, for study $\ell$ the non-conformity score for a candidate $y$ is $|y-\hat{f}_\ell(X)|$, and $\hat{f}_\ell$ is one of the following models: neural network, random forest, LASSO and linear regression. 200 pairs of data are randomly picked as training data, and the rest are used as calibration data. All studies use the same split.
		We use the package \textsf{conformalInference}\footnote{Code is provided in \url{https://github.com/ryantibs/conformal}.} to implement these methods. 
		Since there are only four learning algorithms, we have $L=4$ in this experiment and Scenario 2 is omitted. The result is summarized in Figure \ref{fig:sim_dep_mmthd}. Similar to Section \ref{sec:simu1}, all variations of SAT achieve the target coverage rate for the merged set. We observe that \texttt{SyP+R\"uger} slightly outperforms \texttt{SyE+AM} in some cases. Again this does not contradict the admissibility theory, as \texttt{SyP+R\"uger} is a randomized merging procedure. In contrast, both \texttt{SyP(naïve)+AM} and \texttt{SyP(naïve)+R\"uger} are deterministic, and the resulting sizes from these two procedures are consistently larger than those of \texttt{SyE+AM}. 

\begin{figure}[ht]
			\centering
			\includegraphics[width=\linewidth]{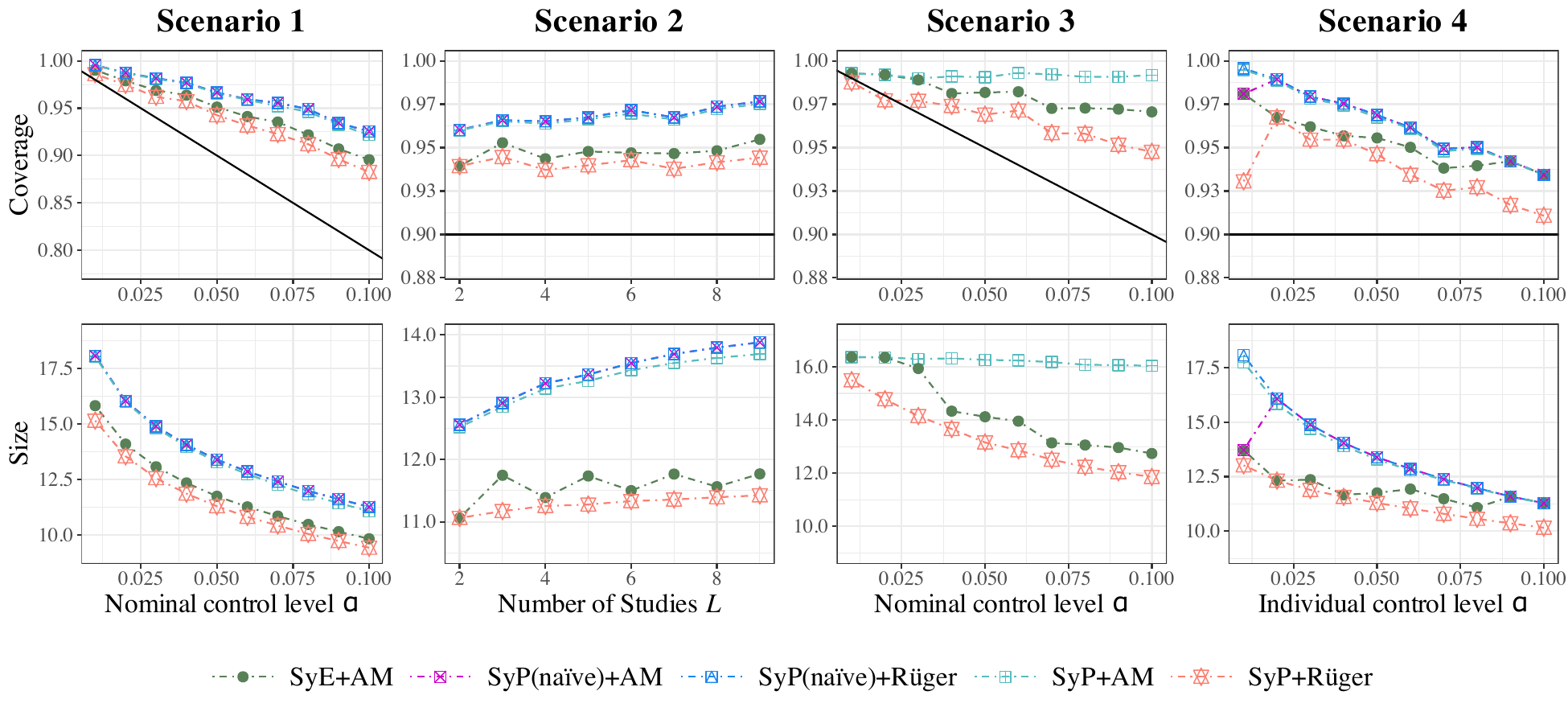}
			\caption[Split-conformal LASSO]{\small Coverage and size of the merged conformal prediction sets by different splits of the training and calibration data. The initial sets are constructed using a split conformal approach with LASSO selected as the score function.}
			\label{fig:sim_dep_lasso}
		\end{figure}

		\noindent\textbf{Conformal prediction set with different splits of training and calibration data.}
		In this experiment, we merge split conformal prediction sets that are constructed using the same learning algorithm but different splits of the training and calibration data. The non-conformity for a candidate $y$ is $|y-\hat{f}_\ell(X)|$ and $\hat{f}_\ell$ is obtained from a LASSO model. Each study randomly selects 200 data points from the 400 labeled samples  for training and uses the remaining 200 points for calibration, denoted as $D^{(\ell)}_{\rm tr}$ and $D^{(\ell)}_{\rm cal}$, respectively. 
		The results are summarized in Figure \ref{fig:sim_dep_lasso}. It shows that all variations of SAT successfully achieve the target coverage level, with \texttt{SyE+AM} performing comparably to \texttt{SyP+R\"uger}, and both uniformly outperforming \texttt{SyP(naïve)+AM} and \texttt{SyP(naïve)+R\"uger}.
		
		
		\section{Real Data Analysis}\label{sec:real}
		We evaluate the performance of the proposed methods on the \textsf{ImageNet\_val} dataset \citep{deng2009imagenet}. The data contains 50,000 labeled images across 1,000 distinct classes.
		Our objective is to merge the prediction sets for class labels generated by various learning algorithms while maintaining a high coverage rate. 
		For instance, when an image of a fox squirrel is provided, different algorithms might yield distinct prediction sets, such as $\cC_1=\{\text{fox squirrel, gray fox, bucket, rain barrel}\}$, $\cC_2=\{\text{marmot, fox squirrel, mink}\}$, etc; our proposed methods will then be employed to aggregate these sets.
		To construct the initial prediction sets, we utilize \texttt{RAPS}, a modified conformal prediction algorithm introduced by \cite{angelopoulos2020uncertainty}. 
		The learning algorithms employed by different studies are \textsf{VGG16}, \textsf{DenseNet161}, \textsf{ResNeXt101}, \textsf{ResNet50}, and \textsf{ResNet18}. 
		We utilize the pre-trained versions of these models, meaning that only calibration data is needed to construct the prediction sets.
		For each replication, we apply stratified random splitting to divide the dataset into five calibration sets of sample size 8,000 and one test set of sample size 10,000, with each study accessing a distinct calibration set.  
		Given that the number of studies is fixed at 5, we generate the initial sets according to Scenarios 1, 3, and 4 as described in Section \ref{sec:simulation}. 
		We replicate the experiments 20 times.
		Note that, \texttt{RAPS} does not explicitly produce ``oracle p-values'', so we omit the comparisons to \texttt{OrP}. 
		Additionally, since \texttt{RAPS} treats the classes of images as random and ensures marginal coverage, only dependent aggregation methods are valid and thus employed in this analysis.
		
		\begin{figure}[t]
			\centering
			\includegraphics[width=\linewidth]{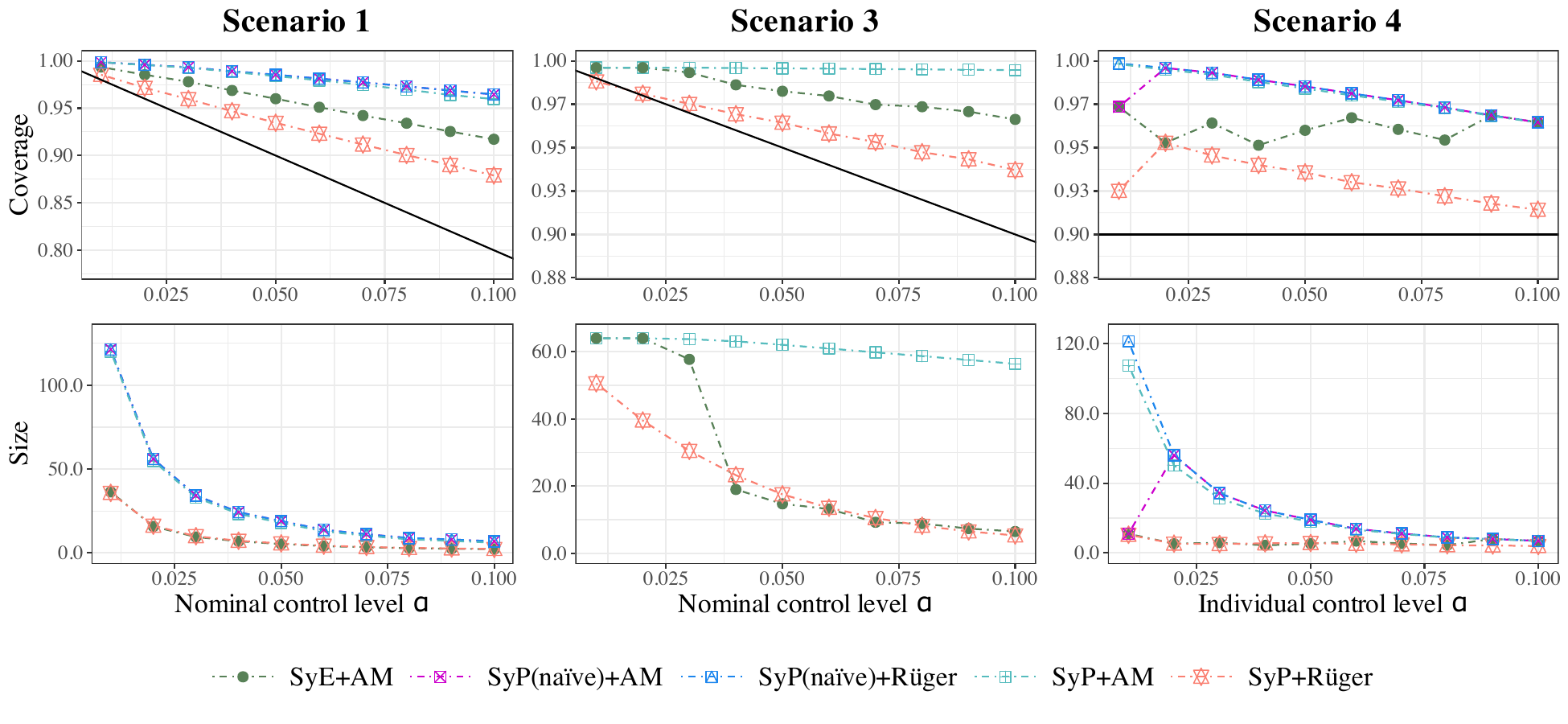}
			\caption[ImageNet results]{Coverage and size of the merged prediction sets using different learning algorithms for dataset \textsf{ImageNet\_val}. }
			\label{fig:rd_imgnet}
		\end{figure}
		The results are summarized in Figure \ref{fig:rd_imgnet}. 
		In comparison to the simulated data from the first part of Section \ref{sec:dependent_simu}, which considers similar setups in linear models, the image data and initial set constructions in the current real data context are significantly more complex.
		Nonetheless, since SAT relies solely on the generated initial sets, its overall performance is similar to that shown in Figure \ref{fig:sim_dep_mmthd}, demonstrating the effectiveness and robustness of the proposed procedures.

		\section{Discussions}\label{sec:dis}
		In this paper, we introduced the SAT framework for merging uncertainty sets in settings where only the initial uncertainty sets and their corresponding control levels are available.
		The proposed method is flexible, computationally efficient, and requires minimal information from each individual study.
		
		The size of the merged set produced by SAT critically depends on the aggregation method used to combine synthetic statistics. As shown in Theorem \ref{admissibility2}, 
		all admissible deterministic set merging procedures under general dependence must be in the form of SAT. 
		In contrast to the clear characterization of deterministic procedures, the notion of admissibility for randomized set merging is less straightforward. Our simulation studies suggest that randomized procedures can produce smaller sets, but defining their optimality is challenging. A formal comparison requires a coherent way to order random sets, for which there is no single canonical choice. One might consider theoretical criteria such as stochastic or almost-sure set inclusion. Alternatively, a more practical ordering could be based on the expected size of the merged sets, though this criterion would depend on the choice of an appropriate measure on the candidate space. A rigorous framework for admissibility in randomized settings, together with a deeper understanding of how various optimality criteria relate to one another, represents a promising avenue for future investigation.

		
		

\bibliographystyle{apalike}
\bibliography{reference}

	\newpage

	\begin{center}
		{\Large Supplementary Material for ``Data-light Uncertainty Set Merging with Admissibility"}
	\end{center}
	\medskip
	
	\begin{center}
	{ Shenghao Qin, Jianliang He, Qi Kuang, Bowen Gang, and Yin Xia}
	\end{center}
\bigskip
\bigskip
	\numberwithin{equation}{section}
	\numberwithin{figure}{section}
	\numberwithin{theorem}{section}
	\numberwithin{definition}{section}
	\numberwithin{assumption}{section}
	\numberwithin{proposition}{section}
	\numberwithin{lemma}{section}
	\numberwithin{remark}{section}

	\appendix
	
	This supplement begins with a content analysis of the merged set under dependence (Section \ref{sec:size_dependence}) and an extension for merging risk-controlling uncertainty sets (Section \ref{ap:risk_control}). We then provide the proofs for all results in the main text and the Supplement (Section \ref{ap:proof}), followed by further discussions and numerical results (Sections \ref{ap:discuss_thm_dep_power} and \ref{ap:sensitivity}). The supplement concludes with a detailed discussion of the admissible deterministic set merging procedure under independence (Section \ref{ap:merge_indep_no_random}).

	\section{Content Analysis under Dependence}\label{sec:size_dependence}

	In this section, we examine the content of merged set under dependence.
	Note that addressing arbitrary dependence is quite challenging, even under strong assumptions \citep{liu2019cauchy}. 
	Consequently, we focus on the special case where uncertainty sets are constructed from Gaussian summary statistics with weak dependence, as outlined below.
 
		 \begin{definition}[\texorpdfstring{$\delta$-weakly dependent uncertainty set}{delta-weakly dependent uncertainty set}] 
			 Let $\Zb=(Z_1,\dots,Z_L)\sim\cN(\bmu,\Sigma)$ be a vector of summary statistics, where $\bmu\in\RR^L$ and $\Sigma=(\sigma_{ij})\in\RR^{L\times L}$ satisfies $L^{-2}\sum_{i=1}^L\sum_{j=1}^L|\sigma_{ij}|=O(L^{-\delta})$ for some absolute constant $\delta>0$. 
			 For the uncertainty sets $\{\cC_{\ell}\}_{\ell\in[L]}$ for $Y\in\cY$ constructed from $\Zb$, we say that $\{\cC_{\ell}\}_{\ell\in[L]}$ are weakly dependent if the joint probability satisfies 
			 $$
			 \PP(y\in \cC_{1,\alpha_1}, \dots, y\in \cC_{L,\alpha_L}) = \PP\left(\frac{|Z_1-y|}{\sqrt{\sigma_{11}}} \leq z_{1-\alpha_1/2}, \dots, \frac{|Z_L-y|}{\sqrt{\sigma_{LL}}} \leq z_{1-\alpha_L/2}\right),\quad\forall y\in\cY.
			 $$
			 \label{def:weak_dependence_set}
		 \end{definition}
		 \vspace{-20pt}
		 Definition \ref{def:weak_dependence_set} naturally extends the concept of weakly dependent Gaussian random variables \citep{lyons1988strong,fan2012estimating}. Specifically, it characterizes the dependencies among initial uncertainty sets by employing the weakly dependent \emph{summary statistics}. 
		 
		Now, we investigate the content of the merged set constructed using synthetic p-values via \eqref{eq:combine_dep_general}.
		We focus on the set generated by Algorithm \ref{alg:pred} for brevity, noting that the results also extend to Algorithm \ref{alg:pred_imp}, based on the extended definition of synthetic p-values or e-values in Section \ref{ap:thm_practice}.
		We begin with some assumptions on the calibrator functions $f_\ell$'s in  \eqref{eq:combine_dep_general}.
		 \begin{assumption}
		  {Assume that $f_\ell = f$ for all $\ell \in [L]$ and $f : [0,\infty) \mapsto [0,\infty]$ satisfies:}
			 \begin{itemize}[leftmargin=0.6cm]
				 \setlength{\itemsep}{-3pt}
				 \item [(\romannumeral1)] {\citep[Admissibility,][]{vovk2021values,gasparin2024combining} $f$ is decreasing, upper semicontinuous, $f(0)=\infty$ and $\|f\|_{L_1([0,1])}=1$;}
				 \item [(\romannumeral2)] (Boundedness) {$|f|<C_f$ on $(0,\infty)$} for some positive constant $C_f>0$.
			 \end{itemize}
			 \label{as:combine_dep}
		 \end{assumption}
		 Assumption \ref{as:combine_dep} is mild. 
		 The admissibility assumption ensures the method cannot be improved by simply using $f/\|f\|_{L_1([0,1])}$ instead of $f$ in \eqref{eq:combine_dep_general}.
		 The boundedness condition is satisfied by most commonly used calibrators, such as $f(p)=L/k\cdot\ind(p\in(0,k/L))+\infty \ind(p=0)$ in R\"uger's method.
		  
		 \begin{theorem}
			Suppose \eqref{eq:miscoverage} holds with $\alpha_\ell=\alpha$ for all $\ell\in [L]$, and $\{\cC_{\ell,\alpha}\}_{\ell\in[L]}$ are $\delta$-weakly dependent for some $\Zb\sim\cN(\mu^*\one_L,\Sigma)$ and $\sigma_\ell^2=1$ for all $\ell\in[L]$. Let $\bar{p}(y)= \inf \{ \alpha \in (0, 1) : \pb(y) \in R_\alpha \}$ for each $y\in\cY$, where $R_\alpha$ is defined in \eqref{eq:combine_dep_general} with $\lambda_\ell=1/L$ for all $\ell\in[L]$.
  			If $f_\ell$'s satisfy Assumption \ref{as:combine_dep} and $\Var[f\{p_1(y)\}]>0$ for all $y\in\cY$, then for any $\alpha'\in (0,1)$, we have 
  			{$$
			 \lim_{L\rightarrow\infty}\PP(y\notin\bar{\cC}_{\alpha'})\to 1, \quad\forall y\in\{y\in\cY:\Gamma(y,\alpha,\alpha')>1\},
			 $$}
			 where
			 $
			 \Gamma(y,\alpha,\alpha')=\left\{\frac{\alpha'}{\alpha}\int_0^{\alpha/\alpha'}f\left(t\right)\rd t-\frac{\alpha'}{1-\alpha}\int_{\alpha/\alpha'}^1f\left(t\right)\rd t\right\}\cdot\left\{\PP(y\notin \cC_{1,\alpha})-\alpha\right\}+\alpha'.
			 $
			 \label{thm:dependent_power}
		 \end{theorem}
		
		 Theorem \ref{thm:dependent_power} indicates that the merged set converges to a subset of $\{y \in \cY : \Gamma(y, \alpha, \alpha') \leq 1\}$. 
		 In contrast to Theorem \ref{thm:ind_power}, the condition $\Gamma(y, \alpha, \alpha') > 1$ is stricter than $\PP(y \notin \cC_{1, \alpha}) > \alpha$ due to the choice of threshold for $\sum_{\ell=1}^{L} \lambda_\ell \cdot f_\ell(p_\ell/\alpha')$ under arbitrary dependence structures.
		  Tighter thresholds and more powerful aggregation methods can be employed if specific dependence structures are assumed \citep{blier2024improved}.
		Concrete examples of calibrators that satisfy the assumptions of Theorem \ref{thm:dependent_power} are discussed in Section \ref{ap:discuss_thm_dep_power}, where it is also demonstrated that the set $\{y\in\cY: \Gamma(y,\alpha,\alpha') > 1\}$ is not empty for certain calibrators.
		For instance, in the case of normal mean estimation, the set $\{y \in \cY : \Gamma(y, \alpha, \alpha') \leq 1\}$ has a finite size when using these calibrators.
	
	Similarly, the size of the merged set constructed using dependent synthetic e-values is examined in the following theorem.
	\begin{theorem}
	Suppose \eqref{eq:miscoverage} holds with $\alpha_\ell=\alpha$ for all $\ell\in [L]$, and $\{\cC_{\ell,\alpha}\}_{\ell\in[L]}$ are $\delta$-weakly dependent for some $\Zb\sim\cN(\mu^*\one_L,\Sigma)$ and $\sigma_\ell^2=1$ for all $\ell\in[L]$.  Let $\bar{e}(y) = G_e\{\eb(y);1\}$ as defined in \eqref{eq:combine_ind_e} for each $y\in\cY$.
	Then, for any $\alpha'\in (0,1)$, we have
	$$
	\lim_{L\rightarrow\infty}\PP(y\notin\bar{\cC}_{\alpha'})\to1, \quad\forall y\in\left\{y\in\cY:\PP(y\notin \cC_{1,\alpha})>\frac{\alpha\tau}{\alpha'}\right\}.
	$$
	where $\bar{\cC}_{\alpha'}=\{y\in\cY: \bar{e}(y)<\tau/\alpha'\}$ for any fixed $\tau\in(0,1]$.
	\label{thm:dependent_power_ev}
	\end{theorem}
	\section{Uncertainty Set Merging with Risk Control}\label{ap:risk_control}


	Motivated by \cite{angelopoulos2022conformal,gasparin2024merging}, this section explores a broader class of uncertainty sets where the natural notion of error extends beyond just the miscoverage rate.
	Again, only initial uncertainty sets and the corresponding control levels are available.
	We consider the scenario where the focus is on a fixed parameter $\theta^*$; the case for random $Y$  is similar and will be omitted here.
	
	Let $\cL:2^\Theta\times\Theta\mapsto[0,B]$ be a bounded \emph{loss function}  satisfying $\cL(\cC,\theta)=0$ if $\theta\in \cC$, where $\cC$ is an uncertainty set and $B\in(0,\infty)$ is known. We say $\mathcal{C}$ controls risk at level $\beta\in(0,B)$ if
	$
	\EE\{\cL(\cC,\theta^*)\}\leq\beta.
	$
	Note that uncertainty sets with a standard miscoverage guarantee can be obtained by simply choosing  $\cL(\cC,\theta^*)=\ind(\theta^*\notin \cC)$ with $B=1$.
	For another example of risk controlling sets, consider a multi-label prediction problem where the prediction target $Y$ is a subset of the potential labels $\{1,\ldots,m\}$.
	In this case, a reasonable choice of the loss function is the missing rate defined as $\cL(\cC,Y)=1-|\cC \cap Y|/|Y|$. 

	Then, for a set of risk controlling sets, i.e., $\{\cC_{\ell,\beta_\ell}, \ell \in [L]: \EE\{\cL(\cC_{\ell,\beta_\ell},\theta^*)\}\leq\beta_\ell\}$, the goal is to combine them and control the risk at a pre-assigned level $\beta$.
	We first generate synthetic p-values as
	\begin{equation}
		p_{\ell}(\theta)\sim\Unif\left(0,\frac{\beta_\ell}{\tau_\ell}\right)\cdot\ind\{\cL(\cC_{\ell,\beta_\ell},\theta)\geq\tau_\ell\}+\Unif\left(\frac{\beta_\ell}{\tau_\ell},1\right)\cdot\ind\{\cL(\cC_{\ell,\beta_\ell},\theta)<\tau_\ell\},
		\label{eq:syp_risk}
	\end{equation}
	where $\tau_\ell\in(\beta_\ell,B)$ is some pre-determined number. Similarly, the synthetic e-values can be generated as
	\begin{equation}
		e_\ell(\theta)=\beta_\ell^{-1}\cdot\cL(\cC_{\ell,\beta_\ell},\theta),\quad\forall(\ell,\theta)\in[L]\times\Theta.
		\label{eq:sye_risk}
	\end{equation}
	Next, we aggregate the synthetic statistics through some appropriate aggregation functions $G(\cdot)$ as studied in Sections \ref{sec:merge_e} and \ref{sec:merge_p}. 
	Finally, the merged uncertainty set can be obtained by
	\begin{equation}
		\bar{\cC}_\beta=\left\{\theta\in\Theta: G\{\pb(\theta)\}>\frac{\beta}{B}\right\}\text{~or~}\bar{\cC}_\beta=\left\{\theta\in\Theta: G\{\eb(\theta)\}<\frac{B}{\beta}\right\}.
		\label{eq:merged_risk_set}
	\end{equation}
	The validity of the merged set is provided below.
	\begin{proposition}
		Suppose the initial sets $\{\cC_{\ell,\beta_\ell}\}_{\ell\in{L}}$ satisfy $\EE\{\cL(\cC_{\ell,\beta_\ell},\theta^*)\}\leq\beta_\ell$ with $\|\cL\|_\infty\leq B$. 
		Let $p_\ell(\cdot)$ and $e_\ell(\cdot)$  be as defined in \eqref{eq:syp_risk}, the merged uncertainty set in \eqref{eq:merged_risk_set} satisfies $\EE\{\cL(\bar{\cC}_\beta,\theta^*)\}\leq\beta$, for any  $\beta\in(0,B)$.
		\label{prop:merged_risk_set}
	\end{proposition}


		The merging method shares a similar high-level concept with that in \cite{gasparin2024merging} and may therefore suffer from over-conservativeness, as the risk control guarantee (see the proof of Proposition \ref{prop:merged_risk_set}) is obtained via
	$$
	\EE\{\cL(\bar{\cC}_\beta,\theta^*)\}= \EE\{\cL(\bar{\cC}_\beta,\theta^*)\ind(\theta^*\notin\bar{\cC}_\beta)\}\leq\|\cL\|_\infty\cdot\sup_{\theta\in\Theta}\PP(\theta\notin\bar{\cC}_\beta).
	$$
	Thus, there remains significant work to be done for merging risk-controlling sets.

	\section{Proofs} \label{ap:proof}
	This section collects the proofs for all theorems, propositions and lemmas.
	\subsection{Proof of Proposition \ref{thm:sye_pred}}
\begin{proof}
	It follows by $\EE[e_\ell(Y)]=\alpha_\ell^{-1}\cdot\EE[\ind(Y\notin \cC_{\ell})]=\alpha_\ell^{-1}\cdot\PP(Y_{}\notin \cC_{\ell})\leq1$.
\end{proof}

	\subsection{Proof of Proposition \ref{thm:agg_e_ind} }

\begin{proof}
The proof follows from the linearity of expectation and the mutual independence of the e-values. By the definition of $\bar{e}_k(Y)$ in \eqref{eq:combine_ind_e}, we have:
\begin{align*}
	\EE\{\bar{e}_k(Y)\} &= \EE\left\{ \binom{L}{k}^{-1}\sum_{\cI_k \in \mathcal{B}_k} \prod_{\ell\in \cI_k} e_\ell(Y) \right\} \\
	&= \binom{L}{k}^{-1} \sum_{\cI_k \in \mathcal{B}_k} \EE\left\{\prod_{\ell\in \cI_k} e_\ell(Y)\right\} \quad \text{(by linearity of expectation)} \\
	&= \binom{L}{k}^{-1} \sum_{\cI_k \in \mathcal{B}_k} \prod_{\ell\in \cI_k} \EE\{e_\ell(Y)\} \quad \text{(by mutual independence)}.
\end{align*}
Since each $e_\ell(Y)$ is a valid e-value, we are given that $\EE\{e_\ell(Y)\} \le 1$ for all $\ell \in [L]$. Furthermore, as e-values are non-negative random variables, their expectations are also non-negative. This allows us to bound the product term:
$$
\prod_{\ell\in \cI_k} \EE\{e_\ell(Y)\} \le \prod_{\ell\in \cI_k} 1 = 1.
$$
Substituting this inequality back into our expression for the expectation of $\bar{e}_k(Y)$, we obtain:
\begin{align*}
	\EE\{\bar{e}_k(Y)\} &\le \binom{L}{k}^{-1} \sum_{\cI_k \in \mathcal{B}_k} 1.
\end{align*}
The sum is over all possible $k$-element subsets of $[L]$, and the number of such subsets is $|\mathcal{B}_k| = \binom{L}{k}$. Therefore, the sum evaluates to $\binom{L}{k}$. This gives us the final result:
\begin{align*}
	\EE\{\bar{e}_k(Y)\} &\le \binom{L}{k}^{-1} \cdot \binom{L}{k} = 1.
\end{align*}
Thus, $\bar{e}_k(Y)$ is a valid e-value, which completes the proof.
\end{proof}

	\subsection{Proof of Proposition \ref{thm:agg_e_dep} }
	\begin{proof}
		The result follows directly from the linearity of expectation:
		\[ \EE\{\bar{e}(Y)\} = \sum_{\ell=1}^{L}\lambda_\ell \EE\{e_\ell(Y)\} \le \sum_{\ell=1}^{L}\lambda_\ell \cdot 1 = 1.  \]
	\end{proof}

	\subsection{Proof of Propositions  \ref{thm:alg_pred}}
\begin{proof}
Let $r_e(y) = \ind\{e(y) \geq \tau/\alpha\}$. It holds that
	\begin{align*}
	\EE\{r_e(Y_{})\} = \PP\{e(Y_{}) \geq \tau/\alpha\} \leq \alpha\cdot\EE\{e(Y_{})\}/\tau \leq \alpha/\tau.
	\end{align*}
	Thus, for the e-value-based approach with $\cC_\alpha = \{y\in\cY: e(y)<\tau/\alpha\}$, we have
	\begin{align*}
		\PP(Y\in\cC_\alpha) 
		= \EE[\PP\{e(y)<\tau/\alpha|Y=y\}]= \PP\{e(Y)<\tau/\alpha\}= 1- \EE[r_e(Y_{})]\geq 1-\alpha/\tau.
	\end{align*}
\end{proof}

\subsection{Proof of Theorem \ref{thm:main_validity}}

\begin{proof}
	The proof follows by demonstrating the validity of each of the three steps in the SAT procedure (Synthetics, Aggregation, and Test inversion) for both the e-value and p-value constructions.
	
	\textbf{Case 1: Merging with Synthetic e-values.}
	First, for each initial uncertainty set $\cC_\ell$ satisfying the miscoverage guarantee in \eqref{eq:miscoverage}, the synthetic e-function $e_\ell(\cdot)$ defined in \eqref{eq:sye_pred} produces a valid e-value. Specifically, Proposition~\ref{thm:sye_pred} establishes that $\EE\{e_\ell(Y)\} \le 1$ for all $\ell \in [L]$.
	
	Second, the aggregated statistic $\bar{e}(Y)$ is constructed using a valid aggregation function as specified in Proposition~\ref{thm:agg_e_ind} (for independent sets) or Proposition~\ref{thm:agg_e_dep} (for dependent sets). These propositions guarantee that if the input statistics are valid e-values, the aggregated statistic $\bar{e}(Y)$ is also a valid e-value, satisfying $\EE\{\bar{e}(Y)\} \le 1$.
	
	Finally, the merged set $\bar{\cC}_\alpha$ is formed by inverting the aggregated e-value, $\bar{\cC}_\alpha = \{ y \in \cY : \bar{e}(y) < 1/\alpha \}$, where we use the default choice of $\tau=1$. The validity of this test inversion is established in Proposition~\ref{thm:alg_pred}. 
	
	\textbf{Case 2: Merging with Synthetic p-values.}
	First, the synthetic p-value $p_\ell(Y)$ generated by the procedure in \eqref{eq:syp_pred} is valid. Proposition~\ref{thm:syp_pred} shows that $p_\ell(Y)$ is marginally super-uniform, satisfying $\PP\{p_\ell(Y) \le t\} \le t$ for all $t \in [0,1]$ and for each $\ell \in [L]$.
	
	Second, these synthetic p-values are combined using a valid aggregation function from Proposition~\ref{thm:agg_p_dep} (for dependent sets) or Proposition~\ref{thm:agg_p_ind} (for independent sets). These propositions ensure that the aggregated statistic $\bar{p}(Y)$ is also a valid, super-uniform p-value, satisfying $\PP\{\bar{p}(Y) \le t\} \le t$ for all $t \in [0,1]$.
	
	Finally, the merged set is constructed via test inversion: $\bar{\cC}_\alpha = \{ y \in \cY : \bar{p}(y) > \alpha \}$. The validity of this step is given in Proposition~\ref{thm:alg_pred_p}. 
	
	Since the procedure guarantees $(1-\alpha)$-coverage for both the e-value and p-value constructions, the theorem is proved.
\end{proof}

\subsection{Proof of Theorem \ref{admissibility}}\label{sec:pfthm4}
\begin{proof}

	We first introduce some terminologies. A function $e:\mathcal{Y}\mapsto\{0,1/\alpha\}$ is called an \textit{e-function}.
	Let $\cC$ be a level $\alpha$ uncertainty set for $Y$. Define its corresponding e-function as $\xi_\alpha(\cC)(y):=\frac{1}{\alpha}\ind(y \notin \cC)$. Note that $\xi_\alpha(\cC)(y)$ is precisely the synthetic e-value construction in \eqref{eq:sye_pred}.

	Conversely, if we have an e-function $e$, define $\psi_\alpha(e):= \left\{ y \in \mathcal{Y} : e(y) < \tfrac{1}{\alpha} \right\},$ then $\psi_\alpha(\cdot)$ is the inverse function of $\xi_\alpha(\cdot)$.
	Denote $\mathcal{E}_{\alpha_\ell}$ the set of all e-functions that maps $\mathcal{Y}$ to $\{0,1/\alpha_\ell\}$.

	We call $F$ an e-function merger if $F: \prod_{\ell=1}^L \mathcal{E}_{\alpha_\ell}\mapsto \mathcal{E}_{\alpha}$, it is \textit{valid} if $\mathbb{E}[F(e_1, \ldots, e_L)(Y)] \leq 1$.
	A valid e-function merger $F$ is called \textit{admissible} if $G$ is another valid e-function merger and $G(e_1,\ldots, e_L)(y)\geq F(e_1,\ldots,e_L)(y)$ for all ($e_1,\ldots, e_L) \in \prod_{\ell=1}^L \mathcal{E}_{\alpha_\ell}$ and $y\in \mathcal{Y}$ the we must have $G=F$.

	Let $f$ be a level $\alpha$ set merging function as defined in Definition \ref{func:setmerge}. Define its corresponding e-function merger as $F(e_1,\ldots,e_\ell)   := \xi_{\alpha} \circ f \circ (\psi_{\alpha_1}(e_1), \dots, \psi_{\alpha_L}(e_L))$.
	By Proposition \ref{thm:sye_pred} we have $\mathbb{E}(F(e_1,\ldots,e_L)(Y))\leq 1$.
	
	Conversely, if we have a valid e-function merger $F$, then it induces a set merging function $f(\cC_1,\ldots,\cC_L):=\psi_{\alpha}\circ F\circ (\xi_{\alpha_1}(\cC_1),\ldots, \xi_{\alpha_\ell}(\cC_\ell))$. By Proposition \ref{thm:alg_pred}, $f$ is a level $\alpha$ set merging function. Therefore, there is a one to one correspondence between valid e-function merger and level $\alpha$ set merging function. Moreover, $f$ is admissible if and only if its corresponding e-function merger $F$ is admissible. We state a useful lemma.

	\begin{lemma}\label{lemma1}
		Let $F$ be an valid e-function merger. There exists an valid e-function merger $G$ such that $G(e_1,\ldots,e_L)(y)$ depends solely on $e_1(y), \ldots, e_L(y) $ and $G(e_1,\ldots,e_L)(y)\geq F(e_1,\ldots,e_L)(y)$. Moreover, if $F$ is symmetric, that is $F(e_1,\ldots,e_L)(y)=F(e_{\sigma(1)},\ldots,e_{\sigma(L)})(y) $, for any permutation $\sigma$ of $\{1,\ldots, L\}$, then $G$ is also symmetric. We call such $G$ a local e-function merger.
	\end{lemma}
	
	The local e-function merger in Lemma \ref{lemma1} induces a function $\tilde{G}:\prod_{\ell=1}^{L}\{0,\frac{1}{\alpha_\ell}\}\mapsto \{0,\frac{1}{\alpha}\}$ such that $\tilde{G}(e_1(y),\ldots,e_L(y))=G(e_1,\ldots,e_L)(y)$. 
	The precise definition of $\tilde{G}$ is as follows. Given $(x_1,\ldots,x_L)\in \prod_{\ell=1}^{L}\{0,\frac{1}{\alpha_\ell}\} $, let $C_1,\ldots, C_L$ and $y$ be such that $x_\ell=\frac{1}{\alpha_\ell}\ind(y\notin C_\ell)$ for $\ell \in[L]$ and let $e_\ell(\cdot)$ be the e-function defined by $(C_\ell,\alpha_\ell)$, then $\tilde{G}(x_1,\ldots,x_L)=G(e_1,\ldots,e_L)(y)$. Note that such $\tilde{G}$ is well defined because $G$ is a local e-function merger so $G(e_1,\ldots,e_L)(y)$ depends only on $e_1(y),\ldots, e_L(y)$ and $e_1(y),\ldots, e_L(y)$ depend only on $x_1,\ldots,x_L$ by construction. 
	We call such $\tilde{G}$ a synthetic e-merging function. Note that by the validity of $G$ we have  $\mathbb{E}(\tilde{G}(e_1(Y),\ldots,e_L(Y)))\leq 1$. 
	Conversely,  a synthetic e-merging function $\tilde{G}$ that satisfies $\EE(G(X_1,\ldots,X_L))\leq 1$ whenever $\EE(X_\ell)\leq1$ for all $\ell\in [L]$ 	   induces a local e-function merger $G$ by defining $G(e_1,\ldots,e_L)(y)=\tilde{G}(e_1(y),\ldots, e_L(y))$.

	Write $x=(x_1,\ldots, x_L)$, we say $\tilde{G}$ is a \textit{convex synthetic e-merging} function if $\tilde{G}(x) = \tfrac{1}{\alpha}\ind\big(\lambda \cdot x \geq \tfrac{1}{\alpha}\big)$ for some $\lambda \in \Delta^{L-1} := \left\{ \lambda \in [0,1]^L : \mathbf{1}^\top \lambda = 1 \right\}$. A synthetic e-merging funcion $\tilde{G}$ is called \textit{admissible} if there is no synthetic e-merging funcion $H\neq \tilde G$ such that $H(\cdot) \geq \tilde{G}(\cdot)$ for all valid input.
	The next lemma characterizes admissible synthetic e-merging function.
	
	\begin{lemma}\label{optimality}
		An admissible synthetic e-merging functions must be a convex synthetic e-merging function.\end{lemma}
	We claim Lemma \ref{lemma1} and Lemma \ref{optimality} together with the duality of set merging function and e-function merger implies Theorem \ref{admissibility}.
	To see this, given a deterministic set merging function $f_1$,  let \( F_1 \) be the  e-function merger induced by \( f_1 \). By Lemma \ref{lemma1}, there exists a local e-function merger $G_{F_1} \geq F_1$, $G_{F_1}$ further induces a synthetic e-merging function $\tilde{G}_{F_1}$. By Lemma \ref{optimality}, we know that there exists a convex synthetic e-merging function \( \tilde{F}_2 \geq \tilde{G}_{F_1} \).
	
	Denote \( F_2 \) the local e-function merger induced by $\tilde{F}_2$ and $f_2$ the deterministic set merging function induced by $F_2$, we have
	\begin{align*}
		f_2(\cC_1, \dots, \cC_L) &= \psi_{\alpha} \circ F_2 \left( \xi_{\alpha_1}(\cC_1), \dots, \xi_{\alpha_L}(\cC_L) \right) \\
		&\subseteq \psi_{\alpha} \circ {G}_{F_1} \left( \xi_{\alpha_1}(\cC_1), \dots, \xi_{\alpha_L}(\cC_L) \right) \\
		&\subseteq \psi_{\alpha} \circ F_1 \left( \xi_{\alpha_1}(\cC_1), \dots, \xi_{\alpha_L}(\cC_L) \right) \\
		&= \psi_{\alpha} \circ \xi_{\alpha} \circ f_1 \left( \psi_{\alpha_1} \circ\xi_{\alpha_1}(\cC_1), \dots, \psi_{\alpha_L} \circ \xi_{\alpha_L}(\cC_L) \right) \\
		&= f_1 \left( \cC_1, \dots, \cC_L \right). \\
	\end{align*}
	By construction $f_2$ is equivalent to SAT with synthetic e-value and convex combination in the aggregation step.
\end{proof}

	\subsection{Proof of Theorem  \ref{admissibility2}}
\begin{proof}
	We first state a lemma.
	\begin{lemma}\label{symmetric}
		
		\( F(x):= \frac{1}{\alpha} \ind\left(\frac{1}{L}\sum_{\ell=1}^{L}x_\ell\geq \frac{1}{\alpha}\right) \) is the only admissible synthetic e-merging function among all valid synthetic e-merging function that maps $\{0,1/\alpha_1\}^L$ to $\{0,1/\alpha\}$.

	\end{lemma}
	$F$ naturally induces a local e-function merger $\hat{F}$. 	Denote the deterministic set merging function induced by \( \hat{F} \) as \( f \). Then 	  	$
	f(\cC_1, \dots, \cC_L) = \psi_{\alpha} \circ \hat{F}(\xi_{\alpha_1}(\cC_1), \dots, \xi_{\alpha_1}(\cC_L)).
	$ Let $h$ be a symmetric deterministic set merging function, we show that $f \subseteq h$.
	
	Let 	  	\( H \) be the  e-function merger induced by \( h \):
	\[
	H(e_1, \dots, e_L) = \xi_{\alpha} \circ h(\psi_{\alpha_1}(e_1), \dots, \psi_{\alpha_1}(e_L)).
	\]
	$H$ is symmetric by definition. By Lemma  \ref{lemma1},
	there exists symmetric local e-merger function  $G_H \geq H$. By Lemma \ref{symmetric}, $F\geq \tilde{G}_H$, which is the synthetic e-merging function induced by $G_H$. Hence, $\hat{F}\geq G_H$.
	The symmetry of \( F \) implies the symmetry of \( f \). We have
	\begin{align*}
		f(\cC_1, \dots, \cC_L) &= \psi_{\alpha} \circ \hat{F} \left( \xi_{\alpha_1}(\cC_1), \dots, \xi_{\alpha_1}(\cC_L) \right) \\
		&\subset \psi_{\alpha} \circ G_H \left( \xi_{\alpha_1}(\cC_1), \dots, \xi_{\alpha_1}(\cC_L) \right) \\
		&\subset \psi_{\alpha} \circ H \left( \xi_{\alpha_1}(\cC_1), \dots, \xi_{\alpha_1}(\cC_L) \right) \\
		&= \psi_{\alpha} \circ \xi_{\alpha} \circ h \left( \psi_{\alpha_1} \circ \xi_{\alpha_1}(\cC_1), \dots, \psi_{\alpha_1} \circ \xi_{\alpha_1}(\cC_L) \right) \\
		&= h \left( \cC_1, \dots, \cC_L \right). \\
	\end{align*}
\end{proof}

	\subsection{Proof of Proposition  \ref{thm:syp_pred}}
\begin{proof}
	By the definition of synthetic p-values in \eqref{eq:syp_pred}, we have
	\begin{align}
		&\PP(p_\ell(Y)\leq t)\notag\\
		&\quad=\PP(p_\ell(Y)\leq t\mid Y\notin \cC_{\ell})\cdot\PP(Y\notin \cC_{\ell})+\PP(p_\ell(Y)\leq t\mid Y\in \cC_{\ell})\cdot\PP(Y\in \cC_{\ell})\notag\\
		&\quad=\frac{t}{\alpha_\ell}\cdot\ind\left(t\leq\alpha_\ell\right)\cdot\PP(Y\notin \cC_{\ell})+\ind\left(t>\alpha_\ell\right)\cdot\PP(Y\notin \cC_{\ell})+\frac{t-\alpha_\ell}{1-\alpha_\ell}\cdot\ind\left(t>\alpha_\ell\right)\cdot\PP(Y\in \cC_{\ell})\notag\\
		&\quad=t-t\left(1-\frac{\PP(Y\notin \cC_{\ell})}{\alpha_\ell}\right)\notag\\
		&\qquad+\underbrace{\left(1-\frac{t}{\alpha_\ell}\right)\cdot\ind\left(t>\alpha_\ell\right)\cdot\PP(Y\notin \cC_{\ell})}_{\textbf{(\romannumeral1)}}+\underbrace{\frac{t-\alpha_\ell}{1-\alpha_\ell}\cdot\ind\left(t>\alpha_\ell\right)\cdot\PP(Y\in \cC_{\ell})}_{\textbf{(\romannumeral2)}}.
		\label{eq:spv-decom}
	\end{align}
	Simple calculation yields that
	\begin{align}
		\textbf{\small(\romannumeral1)}+\textbf{\small(\romannumeral2)}
		=&\frac{t-\alpha_\ell}{1-\alpha_\ell}\cdot\ind\left(t>\alpha_\ell\right)\cdot\PP(Y\in \cC_{\ell})+\left(1-\frac{t}{\alpha_\ell}\right)\cdot\ind\left(t>\alpha_\ell\right)\cdot\PP(Y\notin \cC_{\ell})\notag\\
		=&\frac{t-\alpha_\ell}{1-\alpha_\ell}\cdot\ind\left(t>\alpha_\ell\right)\left(\PP(Y\in \cC_{\ell}) - \frac{1-\alpha_\ell}{\alpha_\ell}\cdot\PP(Y\notin \cC_{\ell})\right)\notag\\
		=&\frac{t-\alpha_\ell}{1-\alpha_\ell}\cdot\ind\left(t>\alpha_\ell\right)\left(1-\frac{\PP(Y\notin \cC_{\ell})}{\alpha_\ell}\right).
		\label{eq:spv-decom-34}
	\end{align}
	Then, combining \eqref{eq:spv-decom} and \eqref{eq:spv-decom-34}, it holds that
	\begin{align*}
		\PP(p(Y)\leq t)&=t+\left(1-\frac{\PP(Y\notin \cC_{\ell})}{\alpha_\ell}\right)\cdot\left(\frac{t-\alpha_\ell}{1-\alpha_\ell}\cdot\ind\left(t>\alpha_\ell\right)-t\right)\\
		&=t+\frac{1}{1-\alpha_\ell}\left(1-\frac{\PP(Y\notin \cC_{\ell})}{\alpha_\ell}\right)\cdot\{(t-\alpha_\ell)\ind\left(t>\alpha_\ell\right)-(t-t\alpha_\ell)\}\leq t,
	\end{align*}
	where the last inequality results from $\PP(Y\notin \cC_{\ell})\leq\alpha_\ell$ and the fact that $(t-\alpha_\ell)\ind\left(t>\alpha_\ell\right)-(t-t\alpha_\ell)<0$ for all $t\in[0,1]$. 
\end{proof}

\subsection{Proof of Proposition \ref{thm:agg_p_dep}}
\begin{proof}
	It follows the same proof as the proof for Theorem 5.1 in \cite{vovk2022admissible}.
\end{proof}

\subsection{Proof of Proposition \ref{thm:agg_p_ind}}
\begin{proof}
Proposition \ref{thm:agg_p_ind} follows by noting that
\begin{align*}
	\PP(\bar p(Y) \leq t) 
	&=\PP(\inf \{ \alpha \in (0, 1) : \mathbf{p}(Y) \in R_\alpha \} \leq t)\\
	&\leq \PP(\pb(Y) \in R_t)= \PP\left(\sum_{\ell=1}^L S_\ell(p_\ell(Y)) \geq c_{1-t}(\{S_\ell\}_{\ell\in[L]})\right)\leq t.
\end{align*}
The first inequality results from the fact that $R_\alpha$ is increasing in $\alpha$ by definition. The last inequality is because $c_{1-t}(\{S_\ell\}_{\ell\in[L]})$ is no smaller than the $(1-t)$-quantile of $\sum_{\ell=1}^L S_\ell(p_\ell(Y))$, as $S_\ell(p_\ell(Y))$ is stochastically no larger than $S_\ell(U_\ell)$ by Proposition \ref{thm:syp_pred} and $S_\ell$'s are decreasing.
\end{proof}
	\subsection{Proof of Proposition  \ref{thm:alg_pred_p} }
\begin{proof}
	By letting $t=\alpha$, we have $\PP\{p(Y_{})\leq \alpha\} \leq \alpha$. Next, let $r_p(y) = \ind\{p(y)\leq\alpha\}$. It holds that
	\begin{align*}
		\EE\{r_p(Y_{})\} = \PP\{p(Y_{})\leq\alpha\} \leq \alpha.
	\end{align*}
	Thus, for the p-value-based approach with $\cC_\alpha = \{y\in\cY: p(y)>\alpha\}$, we have
	\begin{align*}
		\PP(Y\in\cC_\alpha) 
		= \EE[\PP\{p(y)>\alpha|Y=y\}]
		&= \PP\{p(Y)>\alpha\}= 1- \EE[r_p(Y_{})]\geq 1-\alpha,
	\end{align*}

\end{proof}

\subsection{Proof of Theorem \ref{thm:practice}}\label{ap:thm_practice}
\begin{proof}
	
	For any $\tilde{\cY}\in\cM_L$, we define 
	$p_\ell(\tilde{\cY}):=p_\ell(y)$ and $e_\ell(\tilde{\cY}):=e_\ell(y),$
	where $y \in \tilde{\cY}$ is the representative candidate used for constructing synthetic statistics in Algorithm \ref{alg:pred_imp}. Using the notations from Algorithm \ref{alg:pred_imp}, we extend the definition of synthetic statistics to all $y'\in\cY$ and define the mappings $p'_\ell(\cdot)$ and $e'_\ell(\cdot)$ by
	\begin{align}
		\left\{
		\begin{aligned}
			&p'_\ell(y') = \sum_{\tilde{\cY}\in\cM_L} p_\ell(\tilde{\cY})\cdot\ind(y'\in\tilde{\cY}),\\ 
			&e'_\ell(y') = \sum_{\tilde{\cY}\in\cM_L} e_\ell(\tilde{\cY})\cdot\ind(y'\in\tilde{\cY}),
		\end{aligned}\right.\qquad\forall (y',\ell)\in\cY\times[L].
		\label{eq:equiv_syn_alg2}
	\end{align}
	Let $\bar{p}'(y') = G_p(\pb'(y'))$ and $\bar{e}'(y') = G_e(\eb'(y');k)$, where $\pb'(y')=\{p'_1(y'),\ldots,p'_L(y')\}$ and $\eb'(y')=\{e'_1(y'),\ldots,e'_L(y')\}$ for any $y'\in\cY$ and $k\in[L]$. 
	We next claim that Propositions \ref{thm:sye_pred} and \ref{thm:syp_pred} continue to hold when $p_\ell(\cdot)$ and $e_\ell(\cdot)$ are replaced by $p'_\ell(\cdot)$ and $e'_\ell(\cdot)$, respectively.
	Note that for any given $y'\in\cY$, we have $y'\in\tilde{\cY}$ for some $\tilde{\cY}\in\cM_L$. Suppose for this specific $\tilde{\cY}$, we select $y\in\tilde{\cY}$ to construct synthetic statistics in Algorithm \ref{alg:pred_imp}. Then for any $y'\in\cY$ and any $\ell\in[L]$, we have 
	\begin{align*}
		p'_\ell(y')
		&\sim\Unif\left(0,\alpha_\ell\right)\cdot\ind(y\notin\cC_{\ell})+\Unif\left(\alpha_\ell,1\right)\cdot\ind(y\in\cC_{\ell})\\
		&\overset{d}{=}\Unif\left(0,\alpha_\ell\right)\cdot\ind(y'\notin\cC_{\ell})+\Unif\left(\alpha_\ell,1\right)\cdot\ind(y'\in\cC_{\ell}),
	\end{align*}
	where $\overset{d}{=}$ denotes that both sides have the same distribution. The last line follows from the fact $\ind(y\notin\cC_{\ell}) = \ind(y'\notin\cC_{\ell})$ for $y'$ and $y$ in the same $\tilde{\cY}$ for any $\ell\in[L]$, which is guaranteed by the iterative division.
	With a similar argument, we have $e_\ell'(y')=\alpha_\ell^{-1}\cdot\ind(y'\notin\cC_{\ell})$ for any $y'\in\cY$ and $\ell\in[L]$. 
	Then, by Propositions \ref{thm:sye_pred} and \ref{thm:syp_pred}, we have $\EE\{e'_\ell(Y_{})\}\leq1$ for any $\ell\in[L]$ and  $\PP\{p'_\ell(Y_{})\leq t\} \leq t, \forall t\in[0,1]$. 
	It is straightforward to verify that $\cC_{p'} = \{y\in\cY: \bar{p}'(y) > \alpha\}$ and $\cC_{e'} = \{y\in\cY: \bar{e}_k'(y) < \tau/\alpha\}$ are exactly the same as the output of Algorithm \ref{alg:pred_imp} with either synthetic p-values or synthetic e-values.		  	
	The Theorem then follows directly from Propositions \ref{thm:sye_pred} - \ref{thm:alg_pred_p}.

	
\end{proof}

	\subsection{Proof of Theorem \ref{thm:ind_power}}\label{ap:ind_power}
	\begin{proof}
		For any fixed $y\in\cY$ we write $p_\ell(y)=p_\ell$ to simplify notation.
			 Recall that for all $\ell\in[L]$ we have $\alpha_\ell=\alpha$, and
		$p_{\ell}\sim\Unif\left(0,\alpha\right)\cdot\ind(y \notin \cC_{\ell, \alpha})+\Unif\left(\alpha,1\right)\cdot\ind(y \in \cC_{\ell, \alpha})$ as constructed in \eqref{eq:syp_pred}.
		The rejection region for the combination test is of the form
		$$
		R_{\alpha'}=\left\{\pb\in[0,1]^L:\sum_{\ell=1}^LS(p_\ell)\geq c_{1-\alpha'}(S)\right\},
		$$
		where $c_{1-\alpha'}(S)={\rm Quantile}\big(1-\alpha';\sum_{\ell=1}^LS(U_\ell)\big)\text{~with~} U_\ell\iid\Unif(0,1)$.
		Note that Assumption \ref{as:combine_ind} (i) and $\Var(S(U))>0$ imply that 
		$\sum_{\ell=1}^L\frac{S(U_\ell)-\EE[S(U_1)]}{\sqrt{L{\rm Var}\{S(U_1)\}}} \overset{d}{\rightarrow} \mathcal{N}(0,1)$.
		Based on Lemma 21.2 in \cite{van2000asymptotic}, it holds that
		$$
		{\rm Quantile}\left(1-\alpha';\sum_{\ell=1}^L\frac{S(U_\ell)-\EE[S(U_1)]}{\sqrt{L\cdot{\rm Var}\{S(U_1)\}}}\right) \rightarrow z_{1-\alpha'}.
		$$
		Thus, we have
		$
		c_{1-\alpha'}(S)=L\int_0^1S(t)\rd t+\sqrt{L\cdot{\rm Var}\{S(U_1)\}}\cdot z_{1-\alpha'}+o(\sqrt{L}),
		$
		and the asymptotic rejection rule at level $\alpha'\in(0,1)$ can be equivalently written as
		\begin{align}
			\frac{1}{\sqrt{L}}\sum_{\ell=1}^LS(p_\ell)&\geq \sqrt{L}\cdot\int_0^1S(t)\rd t+ \sqrt{{\rm Var}\{S(U_1)\}}\cdot z_{1-\alpha'}+o\left(1\right).
			\label{eq:equiv_rej}
		\end{align}
		Next, we show that the first two moments of $S(p_\ell)$ are bounded. Denote $U_{1,1}\sim\Unif(0,\alpha)$ and $U_{1,2}\sim\Unif(\alpha,1)$ as two independent random variables. By Assumption \ref{as:combine_ind} (i), we have
		\begin{align*}
			\EE[S(p_1)^2] = \EE[S(U_{1,1})^2\cdot\ind(y\notin \cC_{1,\alpha}) + S(U_{1,2})^2\cdot\ind(y\in \cC_{1,\alpha})] \leq \frac{\|S\|_{L_2([0,1])}^2}{\alpha(1-\alpha)} < \infty.
		\end{align*}
		Thus, by central limit theorem, we have
		\begin{align}
			\frac{1}{\sqrt{L}}\left(\sum_{\ell=1}^LS(p_\ell) - L\cdot\EE[S(p_1)]\right) \overset{d}{\rightarrow} \cN(0,\Var\{S(p_1)\}).
			\label{eq:dep_syp_clt}
		\end{align}
		Combine \eqref{eq:equiv_rej} and \eqref{eq:dep_syp_clt}, we have
		\begin{align}
			\PP(y\notin\bar\cC_{\alpha'})\notag
			&=\PP\left(\sum_{\ell=1}^LS(p_\ell)\geq c_{1-\alpha'}(S)\right)\notag\\
			&=\PP\Biggl\{ \frac{1}{\sqrt{L}}\left(\sum_{\ell=1}^LS(p_\ell) - L\cdot\EE[S(p_1)]\right) \notag\\
			&\qquad\quad\geq\sqrt{L}\cdot \left(\int_0^1S(t)\rd t-\EE[S(p_1)]\right) + \sqrt{{\rm Var}\{S(U_1)\}}\cdot z_{1-\alpha'}+o\left(1\right) \Biggl\}\notag\\
			&=\Phi\left(\sqrt{L}\cdot \left(\frac{\int_0^1S(t)\rd t-\EE[S(p_1)]}{\Var\{S(p_1)\}}\right) + \sqrt{\frac{{\rm Var}\{S(U_1)\}}{\Var\{S(p_1)\}}}\cdot z_{1-\alpha'}+o\left(1\right)\right)
			\label{eq:prob_decom}
		\end{align}
		Moreover, by calculations, we have
		\begin{align}
			\int_0^1S(t)\rd t-\EE[S(p_1)]
			&=\int_0^1S(t)\rd t-\EE[\EE[S(p_1)\wvert \cC_{1, \alpha}]\notag]\\
			&=\int_0^1S(t)\rd t-\EE\left[\int_0^\alpha\frac{S(t)}{\alpha}\rd t\cdot\ind(y \notin \cC_{1, \alpha})+\int_\alpha^1\frac{S(t)}{1-\alpha}\rd t\cdot\ind(y \in \cC_{1, \alpha})\right]\notag\\
			&=-\left(\int_0^\alpha \frac{S(t)}{\alpha}\rd t-\int_\alpha^1\frac{S(t)}{1-\alpha}\rd t\right)\cdot\left(\PP(y \notin \cC_{1, \alpha})-\alpha\right)\leq0,
			\label{eq:exp_spl}
		\end{align}
		where the inequality follows fro Condition (\romannumeral2) in Assumption \ref{as:combine_ind}, and the assumption that $\PP(y \notin \cC_{1, \alpha})>\alpha$.
		Furthermore, we define 
		$$
		C = \left(\int_0^\alpha \frac{S(t)}{\alpha}\rd t-\int_\alpha^1\frac{S(t)}{1-\alpha}\rd t\right)^2\cdot\frac{\left(\PP(y \notin \cC_{1, \alpha})-\alpha\right)^2}{2\Var\{S(p_1)\}},
		$$ 
		based on \eqref{eq:prob_decom} and \eqref{eq:exp_spl}, we have
		\begin{align*}
			\PP(y\notin\bar\cC_{\alpha'})
			&=\Phi\left(-\sqrt{2CL} + \sqrt{\frac{{\rm Var}\{S(U_1)\}}{\Var\{S(p_1)\}}}\cdot z_{1-\alpha'}+o\left(1\right)\right)=1-O\left(\exp({-C L})\right),
		\end{align*}
		where the last equality uses the Gaussian tail bounds. 
		  \end{proof}

			\subsection{Proof of Theorem \ref{thm:ind_power_ev}}\label{ap:ind_power_ev}
		  \begin{proof}
			Let $\xi_k = \text{Cov}\left(e_1(y)\prod_{\ell=2}^{k}e_\ell(y),e_1(y)\prod_{\ell=2}^{k}e_\ell'(y)\right)$, where $e_\ell'(y)$ is an independent copy of $e_\ell(y)$. 
			By \eqref{eq:sye_pred}, we have $\EE\prod_{\ell=1}^{k}e_\ell(y) = \left(\alpha^{-1}\PP(y\notin \cC_{1,\alpha})\right)^k$ and $k^2\xi_k \leq k^2/\alpha^{2k} < \infty$. Then, by Theorem 12.3 in \cite{van2000asymptotic}, for any $y\in\cY$, we have
	\begin{align*}
		\sqrt{L}\left( \bar{e}_k(y) - \EE\prod_{\ell=1}^{k}e_\ell(y) \right) \overset{d}{\rightarrow} \cN(0, k^2\xi_k).
	\end{align*}
	For any $y\in\cY$ satisfying that $\PP(y\notin \cC_{1,\alpha}) > \alpha(\tau/\alpha')^{1/k}$, when $L\to\infty$, let $C = \big(\tau/\alpha' - \left(\alpha^{-1}\PP(y\notin \cC_{1,\alpha})\right)^k\big)^2 / 2k^2{\xi_k}$, and then we have
	\begin{align*}
		\PP(y\notin\bar{\cC}_{\alpha'})
		&=\PP(\bar{e}_k(y) \geq \tau/\alpha')\\
		&=1 - O\left(\Phi\left( \sqrt{L} \cdot \frac{\tau/\alpha' - \left(\alpha^{-1}\PP(y\notin \cC_{1,\alpha})\right)^k }{ k\sqrt{\xi_k}} \right)\right)= 1 - O\left(\exp({-C L})\right),
	\end{align*}

	where the last equality is by standard Gaussian tail bounds.
		  \end{proof}

			\subsection{Proof of Theorem \ref{thm:dependent_power}}\label{ap:dependent_power}	  
	  
			  \begin{proof}
				  Let $p^{\rm or}_\ell(y)=2\Phi(-|Z_\ell-y|)$ and we have that $y\in \cC_{\ell,\alpha}$ is equivalent to $p^{\rm or}_\ell(y) > \alpha$ for any $\ell\in[L]$. Therefore, in the following, we will use $y\in \cC_{\ell,\alpha}$ and $p^{\rm or}_\ell(y) > \alpha$ interchangeably; same for $y\notin \cC_{\ell,\alpha}$ and $p^{\rm or}_\ell(y) \leq \alpha$. For any $y\in\cY$ we write $p_\ell(y)=p_\ell$ and $p_\ell^{\rm or}(y)=p^{\rm or }_\ell$ when there is no ambiguity. Note that due to the generation process of $\{p_\ell\}_{\ell=1}^L$ we have
					\begin{align}
					  \sum_{\ell=1}^{L} \lambda_\ell\cdot f_\ell\left(\frac{p_\ell}{\alpha'}\right)
				  &=\frac{1}{L}\sum_{\ell=1}^L f_\ell\left(\frac{p_\ell}{\alpha'}\right)(\ind(p_\ell^{\rm or}\leq\alpha)+\ind(p_\ell^{\rm or}>\alpha))\notag\\
				  &=\frac{1}{L}\sum_{\ell=1}^L f_\ell\left(\frac{p_\ell}{\alpha'}\right)\ind(p_\ell^{\rm or}\leq\alpha)+\frac{1}{L}\sum_{\ell=1}^L f_\ell\left(\frac{p_\ell}{\alpha'}\right)\ind(p_\ell^{\rm or}>\alpha)\notag\\
				  &=\frac{1}{L}\sum_{\ell=1}^L f_\ell\left(\frac{U_{\ell,1}}{\alpha'}\right) \ind(p_\ell^{\rm or}\leq\alpha)+\frac{1}{L}\sum_{\ell=1}^L f_\ell\left(\frac{U_{\ell,2}}{\alpha'}\right)\ind(p_\ell^{\rm or}>\alpha),
				  \label{eq:equiv_rej_dep}
					\end{align}
				  where $U_{\ell,1} \iid \Unif(0,\alpha)$ and $U_{\ell,2} \iid \Unif(\alpha,1)$ for all $\ell\in[L]$ are independent of all other variables and of each other. 
				  Following this, we want to show that
				  \begin{align}
				  \left\{
					\begin{aligned}
					&\frac{1}{L}\sum_{\ell=1}^L f_\ell\left(\frac{U_{\ell,1}}{\alpha'}\right)\ind(p_\ell^{\rm or}\leq\alpha)\overset{p}{\rightarrow}\EE f_\ell\left(\frac{U_{1,1}}{\alpha'}\right)\cdot\PP(p^{\rm or}_1\leq\alpha),\\ 
				  &\frac{1}{L}\sum_{\ell=1}^L f_\ell\left(\frac{U_{\ell,2}}{\alpha'}\right)\ind(p_\ell^{\rm or}>\alpha)\overset{p}{\rightarrow}\EE f_\ell\left(\frac{U_{1,2}}{\alpha'}\right)\cdot\PP(p^{\rm or}_1>\alpha),
					\end{aligned}\right.
					\label{eq:conv_inp_dep}
					\end{align}
				  both at the rate of $L^{\min\{\delta,1\}/2}$.
				  We first show that
					$
					{\rm Var}\left(\frac{1}{L}\sum_{\ell=1}^L f_\ell\left(\frac{U_{\ell,1}}{\alpha'}\right)\ind(p_\ell^{\rm or}\leq\alpha)\right)=O\left(L^{-\min\{\delta,1\}}\right).
					$
					To begin with, note that the variance can be written as
					\begin{align}
						&{\rm Var}\left(\frac{1}{L}\sum_{\ell=1}^L f_\ell\left(\frac{U_{\ell,1}}{\alpha'}\right)\ind(p_\ell^{\rm or}\leq\alpha)\right)\notag\\
					  &\quad=\frac{1}{L^2}\sum_{\ell=1}^L{\rm Var}\left( f_\ell\left(\frac{U_{\ell,1}}{\alpha'}\right)\ind(p_\ell^{\rm or}\leq\alpha)\right)\notag\\
					  &\qquad+\frac{2}{L^2}\sum_{1\leq\ell<k\leq L}{\rm Cov}\left( f_\ell\left(\frac{U_{\ell,1}}{\alpha'}\right)\ind(p_\ell^{\rm or}\leq\alpha), f_\ell\left(\frac{U_{k,1}}{\alpha'}\right)\ind(p_k^{\rm or}\leq\alpha)\right)\notag\\
					  &\quad\leq\frac{C_f^2}{L^2}\sum_{\ell=1}^L{\rm Var}\left(\ind(p_\ell^{\rm or}\leq\alpha)\right)+\frac{2C_f^2}{L^2}\sum_{1\leq\ell<k\leq L}{\rm Cov}\left(\ind(p_\ell^{\rm or}\leq\alpha),\ind(p_k^{\rm or}\leq\alpha)\right)\notag\\
						&\quad\leq\frac{2C_f^2}{L^2}\sum_{1\leq\ell<k\leq L}{\rm Cov}\left(\ind(p_\ell^{\rm or}\leq\alpha),\ind(p_k^{\rm or}\leq\alpha)\right)+\frac{C_f^2}{4L},
						\label{eq:cov_decom}
					\end{align}
				  where the first inequality is implied by Assumption \ref{as:combine_dep} and the independence of $\{U_{\ell,1}\}_{\ell=1}^L$, $\{U_{\ell,2}\}_{\ell=1}^L$ from all other variables and from each other, and the last inequality is because $\ind(p_\ell^{\rm or}\leq\alpha)\sim\text{Bernoulli}(\PP(p_\ell^{\rm or}\leq\alpha))$.
					Recall that $p^{\rm or}_\ell(y)=2\Phi(-|Z_\ell-y|)$  with $\Zb=(Z_1,\dots,Z_L)\sim\cN(\mu^*\one_L,\Sigma)$ and $\sigma_\ell^2=1$ for all $\ell\in[L]$. Thus for any $1\leq\ell\neq k\leq L$, it follows that
					\begin{align}
						&{\rm Cov}\left(\ind(p_\ell^{\rm or}\leq\alpha),\ind(p_k^{\rm or}\leq\alpha)\right)\notag\\
					  &\quad=\PP(p_\ell^{\rm or}\leq\alpha,p_k^{\rm or}\leq\alpha)-\PP(p_\ell^{\rm or}\leq\alpha)\cdot\PP(p_k^{\rm or}\leq\alpha)\notag\\
						&\quad=\PP\left(|Z_\ell-y|\leq-z_{\sfrac{\alpha}{2}},|Z_k-y|\leq-z_{\sfrac{\alpha}{2}}\right)-\PP\left(|Z_\ell-y|\leq-z_{\sfrac{\alpha}{2}}\right)\cdot\PP\left(|Z_k-y|\leq-z_{\sfrac{\alpha}{2}}\right).
					  \label{eq:cov_porlk}
					\end{align}
					It is straightforward to see that $\PP\left(|Z_\ell-y|\leq-z_{\sfrac{\alpha}{2}}\right)=\PP\left(|Z_k-y|\leq-z_{\sfrac{\alpha}{2}}\right)=\Phi(y-\mu^*-z_{\sfrac{\alpha}{2}})-\Phi(y-\mu^*+z_{\sfrac{\alpha}{2}})$.
					Besides, by writing $Z_\ell,Z_k$ into $Z_\ell=Z_\ell^s-Z^c$ and $Z_k=Z_k^s-Z^c$ with $Z_\ell^s,Z_k^s\iid\cN(\mu^*,1-\sigma_{\ell k})$ and $Z^c\iid\cN(0,\sigma_{\ell k})$, we have that
					\begin{align}
					  &\PP\left(|Z_\ell-y|\leq-z_{\sfrac{\alpha}{2}},|Z_k-y|\leq-z_{\sfrac{\alpha}{2}}\right)\notag\\
					  &\quad=\PP\left(y+z_{\sfrac{\alpha}{2}}\leq Z_\ell^s-Z^c\leq y-z_{\sfrac{\alpha}{2}},y+z_{\sfrac{\alpha}{2}}\leq Z_k^s-Z^c\leq y-z_{\sfrac{\alpha}{2}}\right)\notag\\
					  &\quad=\int_{-\infty}^\infty\PP\left(y+z_{\sfrac{\alpha}{2}}+\sigma_{\ell k}^{1/2}z\leq Z_\ell^s\leq y-z_{\sfrac{\alpha}{2}}+\sigma_{\ell k}^{1/2}z\right)^2\cdot\phi(z)\rd z\notag\\
					  &\quad=\int_{-\infty}^\infty\left[\Phi\left(\frac{\sigma_{\ell k}^{1/2}z+y-\mu^*-z_{\sfrac{\alpha}{2}}}{(1-\sigma_{\ell k})^{1/2}}\right)-\Phi\left(\frac{\sigma_{\ell k}^{1/2}z+y-\mu^*+z_{\sfrac{\alpha}{2}}}{(1-\sigma_{\ell k})^{1/2}}\right)\right]^2\phi(z)\rd z.
					  \label{eq:joint_covariance}
				  \end{align}
					Next, we use Taylor expansion to analyze the joint probability further.  For each $\Phi(\cdot)$, we view it as a function of $\sigma_{\ell k}^{1/2}$. By a similar argument as the in proof of Proposition 2 in \cite{fan2012estimating}, we apply Taylor expansion at $0$ to get
					\begin{align}
						\Phi\left(\frac{\sigma_{\ell k}^{1/2}z+c}{(1-\sigma_{\ell k})^{1/2}}\right)
						=\Phi(c)+\phi(c)\cdot z\sigma_{\ell k}^{1/2}+\frac{\phi(c)c}{2}\cdot(1-z^2)\sigma_{\ell k}+R(\sigma_{\ell k}), 
						\label{eq:taylor} 	
					\end{align}
					where $R(\sigma_{\ell k})$ is the Lagrange residual term in the Taylor's expansion, and $R(\sigma_{\ell k})=f(z)\cdot O\big(\sigma_{\ell k}^{3/2}\big)$ in which $f(z)$ is a polynomial function of $z$ with the highest order as 6. Hence, by substituting \eqref{eq:taylor} back into \eqref{eq:joint_covariance}, we have
					\begin{align}
						&\PP\left(|Z_\ell-y|\leq-z_{\sfrac{\alpha}{2}},|Z_k-y|\leq-z_{\sfrac{\alpha}{2}}\right)\notag\\
						&\quad=[\Phi(y-\mu^*-z_{\sfrac{\alpha}{2}})-\Phi(y-\mu^*+z_{\sfrac{\alpha}{2}})]^2\notag\\
						&\qquad+[\phi(y-\mu^*-z_{\sfrac{\alpha}{2}})-\phi(y-\mu^*+z_{\sfrac{\alpha}{2}})]^2\cdot\sigma_{\ell k}+O\big(\sigma_{\ell k}^{3/2}\big)\notag,
					\end{align}
					where we use the facts that $\int_{-\infty}^\infty \phi(z)\rd z=1$, $\int_{-\infty}^\infty z\phi(z)\rd z=0$, $\int_{-\infty}^\infty z^2\phi(z)\rd z=1$, and the fact that finite moments of standard normal distribution are finite. Following this, we rewrite \eqref{eq:cov_porlk} as 
					\begin{equation}
					  {\rm Cov}\left(\ind(p_\ell^{\rm or}\leq\alpha),\ind(p_k^{\rm or}\leq\alpha)\right)=[\phi(y-\mu^*-z_{\sfrac{\alpha}{2}})-\phi(y-\mu^*+z_{\sfrac{\alpha}{2}})]^2\cdot\sigma_{\ell k}+O\big(\sigma_{\ell k}^{3/2}\big).
					  \label{eq:cov_taylor}
				  \end{equation}
					Write $c_0=[\phi(y-\mu^*-z_{\sfrac{\alpha}{2}})-\phi(y-\mu^*+z_{\sfrac{\alpha}{2}})]^2$ and combine \eqref{eq:cov_decom}, \eqref{eq:cov_taylor}, we have
					\begin{align}
						&{\rm Var}\left(\frac{1}{L}\sum_{\ell=1}^L f_\ell\left(\frac{U_{\ell,1}}{\alpha'}\right)\ind(p_\ell^{\rm or}\leq\alpha)\right)\leq\frac{2C_f^2}{L^2}\sum_{1\leq\ell<k\leq L}\left\{c_0\cdot\sigma_{\ell k}+O\big(\sigma_{\ell k}^{3/2}\big)\right\}+\frac{C_f^2}{4L}\notag\\
						&\quad\leq\frac{2C_f^2}{L^2}\sum_{1\leq\ell<k\leq L}O\big(\sigma_{\ell k}\big)+\frac{C_f^2}{4L}\leq O\left(\frac{1}{L^2}\sum_{\ell=1}^L\sum_{\ell=1}^L|\sigma_{ij}|\right)+\frac{C_f^2}{4L}=O\left(L^{-\min\{\delta,1\}}\right),
					  \label{eq:conv_dep_var}
					\end{align}
					where the second inequality follows from the fact that $\sigma_{\ell k}^{3/2}\leq\sigma_{\ell k}$ since $\sigma_{\ell k}\leq\sigma_\ell\cdot\sigma_k=1$, the third inequality follows from the weak dependence of $\{p^{\rm or}_\ell\}_{\ell=1}^L$, and the last line is because $0<\delta\leq 1$. Thus for any $\epsilon>0$, we have
				  \begin{align}
					  &\PP\left(\left|\frac{1}{L}\sum_{\ell=1}^L f_\ell\left(\frac{U_{\ell,1}}{\alpha'}\right)\ind(p_\ell^{\rm or}\leq\alpha) - \EE \left(f_\ell\left(\frac{U_{\ell,1}}{\alpha'}\right)\right)\PP(p^{\rm or}_1\leq\alpha)\right|\geq\epsilon\right)\notag\\
					  &\quad=\PP\left(\left|\frac{1}{L}\sum_{\ell=1}^L f_\ell\left(\frac{U_{\ell,1}}{\alpha'}\right)\ind(p_\ell^{\rm or}\leq\alpha) - \EE\left(f_\ell\left(\frac{U_{\ell,1}}{\alpha'}\right)\ind(p_1^{\rm or}\leq\alpha)\right)\right|\geq\epsilon\right)\notag\\
					  &\quad=\PP\left(\left|\frac{1}{L}\sum_{\ell=1}^L f_\ell\left(\frac{U_{\ell,1}}{\alpha'}\right)\ind(p_\ell^{\rm or}\leq\alpha) - \frac{1}{L}\sum_{\ell=1}^L \EE\left( f_\ell\left(\frac{U_{\ell,1}}{\alpha'}\right)\ind(p_\ell^{\rm or}\leq\alpha)\right)\right|\geq\epsilon\right)\notag\\
					  &\quad\leq\epsilon^{-2}\cdot{\rm Var}\left(\frac{1}{L}\sum_{\ell=1}^L f_\ell\left(\frac{U_{\ell,1}}{\alpha'}\right)\ind(p_\ell^{\rm or}\leq\alpha)\right)=O\left(L^{-\delta}\right).
					  \label{eq:dep_lln_u11}
				  \end{align}
				  The first equality follows by independence of $U_{\ell,1}$ and $p^{\rm or}_1$, the second equality is due to all $ f_\ell\left(\frac{U_{\ell,1}}{\alpha'}\right)\ind(p_\ell^{\rm or}\leq\alpha)$ having the same distribution, the inequality follows from Chebyshev's inequality, and the last line is derived from \eqref{eq:conv_dep_var}. This completes the proof of the first part of \eqref{eq:conv_inp_dep}.
				  By similar arguments, we can prove the second part.
				  Next, we calculate the expectation 
				  \begin{align}
					  &\EE f_\ell\left(\frac{U_{1,1}}{\alpha'}\right)\cdot\PP(p^{\rm or}_1\leq\alpha) + \EE f_\ell\left(\frac{U_{1,2}}{\alpha'}\right)\cdot\PP(p^{\rm or}_1>\alpha)\notag\\
					  &\quad=\left(\frac{1}{\alpha}\int_0^\alpha f\left(\frac{t}{\alpha'}\right)\rd t-\frac{1}{1-\alpha}\int_\alpha^1 f\left(\frac{t}{\alpha'}\right)\rd t\right)\cdot\left(\PP(p^{\rm or}_1(y)\leq\alpha)-\alpha\right)+\int_0^1 f\left(\frac{t}{\alpha'}\right)\rd t\notag\\
					  &\quad=\left(\frac{\alpha'}{\alpha}\int_0^{\alpha/\alpha'}f\left(t\right)\rd t-\frac{\alpha'}{1-\alpha}\int_{\alpha/\alpha'}^1f\left(t\right)\rd t\right)\cdot\left(\PP(y\notin \cC_{1,\alpha})-\alpha\right)+\alpha':=\Gamma(y,\alpha,\alpha').
					  \label{eq:dep_sum_exps}
				  \end{align}
				  Therefore, based on \eqref{eq:equiv_rej_dep}, \eqref{eq:conv_inp_dep}, \eqref{eq:dep_sum_exps} and Slutsky's theorem, we have	
				  \begin{align*}
					  \lim_{L\rightarrow\infty}\PP(y\notin\bar \cC_{\alpha'})&=\PP\left(\sum_{\ell= 1}^L \lambda_\ell\cdot f_\ell\left(p_\ell,\alpha\right)\geq1\right)=\PP\left(\Gamma(y,\alpha,\alpha') + O_p\left(L^{-\delta/2}\right) \geq 1\right)=1,
				  \end{align*} 
				  for any $y\in\cY$ such that $\Gamma(y,\alpha,\alpha')>1$.
				\end{proof}

			\subsection{Proof of Theorem \ref{thm:dependent_power_ev}}\label{ap:dependent_power_ev}
			 \begin{proof}
				 Recall that $\bar{e}(y) = \sum_{\ell=1}^L e_\ell(y)/L$, and $y\notin\bar \cC_{\alpha'}$ can be rewritten as $\sum_{\ell=1}^L e_\ell(y)/L \geq \tau/\alpha'$, which is equivalent to $
					 \frac{1}{L}\sum_{\ell=1}^L  \ind(y\notin \cC_{\ell,\alpha}) \geq {\alpha\tau}/{\alpha'}$.
				 Mimicking the proof of Theorem \ref{thm:dependent_power}, by taking $f_\ell\left(\frac{U_{\ell,1}}{\alpha'}\right)\equiv1$ in \eqref{eq:dep_lln_u11}, we can obtain that
				 \begin{align*}
					&\PP\left(\left|\frac{1}{L}\sum_{\ell=1}^L \ind(y\notin \cC_{\ell,\alpha}) - \PP(y\notin \cC_{1,\alpha})\right|\geq\epsilon\right)\\
					&\quad=\PP\left(\left|\frac{1}{L}\sum_{\ell=1}^L \ind(p_\ell^{\rm or}\leq\alpha) - \PP(p^{\rm or}_1\leq\alpha)\right|\geq\epsilon\right)
					=O\left(L^{-\delta}\right),
				 \end{align*}
				 which yields 
				 $
				 \sum_{\ell=1}^L \frac{1}{L} \ind(y\notin \cC_{\ell,\alpha}) = \PP(y\notin \cC_{1,\alpha}) + O_p(L^{-\delta/2}).
				 $
 				 Thus, we have
				 \begin{align*}
					 \lim_{L\rightarrow\infty}\PP(y\notin\bar \cC_{\alpha'})&=\PP\left(\PP(y\notin \cC_{1,\alpha}) + O_p(L^{-\delta/2})\geq \frac{\alpha\tau}{\alpha'}\right)=1,
				 \end{align*}  	
				 for any $y\in\cY$ such that $\PP(y\notin \cC_{1,\alpha})>{\alpha\tau}/{\alpha'}$, which completes the proof.
				 \end{proof}

		\subsection{Proof of Proposition \ref{prop:merged_risk_set}}\label{ap:merged_risk_set}
		\begin{proof}
			Following similar arguments in Theorem \ref{thm:syp_pred}, \ref{thm:sye_pred}, we first demonstrate that the synthetic test statistics in \eqref{eq:syp_risk} and \eqref{eq:sye_risk} are valid p-(e-)values. Note that
		\begin{align}
			\PP(p_{\ell}(\theta)\leq t)
			&=\frac{t\tau_\ell }{\beta_\ell}\cdot\ind\left(t\leq\frac{\beta_\ell}{\tau_\ell}\right)\cdot\PP(\cL(\cC_{\ell,\beta_\ell},\theta)\geq\tau_\ell)+\ind\left(t>\frac{\beta_\ell}{\tau_\ell}\right)\cdot\PP(\cL(\cC_{\ell,\beta_\ell},\theta)\geq\tau_\ell)\notag\\
			&\quad+\frac{t-\frac{\beta_\ell}{\tau_\ell}}{1-\frac{\beta_\ell}{\tau_\ell}}\cdot\ind\left(t>\frac{\beta_\ell}{\tau_\ell}\right)\cdot\PP(\cL(\cC_{\ell,\beta_\ell},\theta)<\tau_\ell)\notag\\
			&= t-t\cdot\left(1-\frac{\PP(\cL(\cC_{\ell,\beta_\ell},\theta)\geq\tau_\ell)}{\beta_\ell/\tau_\ell}\right)\notag\\
			&\quad+ \underbrace{\left(1-\frac{t\tau_\ell}{\beta_\ell}\right)\cdot\ind\left(t>\frac{\beta_\ell}{\tau_\ell}\right)\cdot\PP(\cL(\cC_{\ell,\beta_\ell},\theta)\geq\tau_\ell)}_{\textbf{(\romannumeral1)}}\notag\\
			&\quad+\underbrace{\frac{t-\frac{\beta_\ell}{\tau_\ell}}{1-\frac{\beta_\ell}{\tau_\ell}}\cdot\ind\left(t>\frac{\beta_\ell}{\tau_\ell}\right)\cdot\PP(\cL(\cC_{\ell,\beta_\ell},\theta)<\tau_\ell)}_{\textbf{(\romannumeral2)}}.
			\label{eq:decom_risk_1}
		\end{align}
		 Furthermore, following the same calculations in \eqref{eq:spv-decom-34}, we can get
		 \begin{align}
			\textbf{\small(\romannumeral1)}+\textbf{\small(\romannumeral2)}=\frac{t-\frac{\beta_\ell}{\tau_\ell}}{1-\frac{\beta_\ell}{\tau_\ell}}\cdot\ind\left(t>\frac{\beta_\ell}{\tau_\ell}\right)\left(1-\frac{\PP(\cL(\cC_{\ell,\beta_\ell},\theta)\geq\tau_\ell)}{\beta_\ell/\tau_\ell}\right).
			\label{eq:decom_risk_2}
		\end{align}
		Combine \eqref{eq:decom_risk_1} and \eqref{eq:decom_risk_2}, we can get
		$$
		\PP(p_{\ell}(\theta)\leq t)=t+\left(1-\frac{\PP(\cL(\cC_{\ell,\beta_\ell},\theta)\geq\tau_\ell)}{\beta_\ell/\tau_\ell}\right)\cdot\left(\frac{t-\frac{\beta_\ell}{\tau_\ell}}{1-\frac{\beta_\ell}{\tau_\ell}}\cdot\ind\left(t>\frac{\beta_\ell}{\tau_\ell}\right)-t\right)\leq t,
		$$
		where the last inequality results from Markov's inequality and risk control guarantee 
		 $$
		 \frac{\tau_\ell}{\beta_\ell}\cdot\PP(\cL(\cC_{\ell,\beta_\ell},\theta)\geq\tau_\ell)\leq\frac{\EE[\cL(\cC_{\ell,\beta_\ell},\theta)]}{\beta_\ell}\leq1.
		 $$ 
		 Besides, it's easy to show that
		 $\EE[e_\ell(\theta)]=\beta_\ell^{-1}\cdot\EE[\cL(\cC_{\ell,\beta_\ell},\theta)]\leq1$, and thus $e_\ell(\theta)$ is a valid e-value.
		 Based on the arguments above, it is straightforward to show that the merged uncertainty set in \eqref{eq:merged_risk_set} is $\beta$-risk controlled since
		 $$
		 \EE[\cL(\bar{\cC}_\beta,\theta^*)]= \EE[\cL(\bar{\cC}_\beta,\theta^*)\ind(\theta^*\notin\bar{\cC}_\beta)]\leq\|\cL\|_\infty\cdot\sup_{\theta\in\Theta}\PP(\theta\notin\bar{\cC}_\beta)\leq B\cdot\frac{\beta}{B}=\beta,
		 $$
		 where the first equation uses the condition that the loss function satisfies  $\cL(\cC,\theta)=0$ if $\theta\in \cC$.
		\end{proof}

\subsection{Proof of Lemma \ref{lemma1}}

\begin{proof}
	Let us first consider the case where $Y=y$ is fixed, i.e. $F(e_1, \ldots, e_L)(y) = F(e_1(y), \ldots, e_L(y))(y)$.	Suppose $F(e_1, \ldots, e_L)(y)$ depends on $e_i(y), \ldots, e_L(y)$ and 
	$v \in \bigcup_{y^* \neq y, i=1 \ldots L} \{e_i(y^*)\}$. 
	Define $G(e_1, \ldots, e_L)(y) = \sup_{v \in V} F(e_1(y), \ldots, e_L(y), v)(y)$. 	By definition $G$ depends only on $(e_1(y),\ldots,e_L(y))$ and $G \ge F$.
	
		It remains to show $G$ is valid. Let $v_0$ be such that $\sup_v F(e_1(y), \ldots, e_L(y), v)(y)= F(e_1(y), \ldots, e_L(y), v_0)(y)$.
	Such $v_0$ exists because $F(e_1(y), \ldots, e_L(y), v)(y) \in \{0, 1/\alpha\}$.
	Now, define $\tilde{e}_1, \ldots, \tilde{e}_L$ as follows. $\tilde{e}_i(y) = e_i(y)$ for $i=1, \ldots, L$ and $e_i(y) \notin v$.
	If $e_i(y') \in v$, let ${e}^0_i(y')$ be its value in $v_0$. Define $\tilde{e}_i(y) = e^0_i(y')$.
	These e-functions in turn define uncertainty sets $\tilde{\cC}_1, \ldots, \tilde{\cC}_L$. Note that $\mathbb{P}(y \in \tilde{\cC}_i) = \mathbb{P}(y \notin \cC_i)$ since the event $y \in \tilde{\cC}_i$ is solely determined by $\tilde{e}_i(y)$, the event $y \in \cC_i$ is solely determined by $e_i(y)$, and $\tilde{e}_i(y) = e_i(y)$ by construction.
	
	 By definition of $G$ and $\tilde{e}_\ell$, we have $F(\tilde{e}_1, \ldots, \tilde{e}_L)(y) = G(\tilde{e}_1(y), \ldots, \tilde{e}_L(y))(y)$.
	Since $\tilde{\cC}_1, \ldots, \tilde{\cC}_L$ has the same coverage as $\cC_1, \ldots, \cC_L$,  they are also valid uncertainty sets.
	By the validity of $F$, we have $\mathbb{E}[F(\tilde{e}_1, \ldots, \tilde{e}_L)(y)] \le 1$.
	It follows that $\mathbb{E}[G(e_1(y), \ldots, e_L(y))(y)] \le 1$. i.e. $G$ is valid.
	
	If $Y$ is random, we can condition on $Y=y$. Let $h$ be the density function of $Y$.
	$\EE[\ind(Y \in \cC_\ell)] = \int \EE[\ind(Y \in \cC_\ell | Y=y)] h(y) dy$.
	Conditioning on the event $Y=y$ we can construct $\tilde{\cC}_\ell$ and $\tilde{e}_\ell$ the same way as before. Note that $\ind(Y \in \tilde{\cC}_\ell | Y=y) = \ind(Y \in \cC_\ell | Y=y)$.
	Hence if $\EE[\ind(Y \in \cC_\ell)]\geq 1-\alpha_\ell$ then we also have   $\EE[\ind(Y \in \tilde{\cC}_\ell)]\geq 1-\alpha_\ell$. Moreover, by definition of $G$ and $\tilde{e}_\ell$ we have  $G(e_1(y), \ldots, e_L(y))(y) = F(\tilde{e}_1, \ldots, \tilde{e}_L)(y)$. Since $F$ is valid we have
	$\mathbb{E}[F(\tilde{e}_1, \ldots, \tilde{e}_L)(Y)] \le 1$. It follows that $\mathbb{E}[G(e_1(Y), \ldots, e_L(Y))(Y)] \le 1$.
	
	For the second part of the lemma, note that if $F$ is symmetric then $G$ defined by $G(e_1, \ldots, e_L)(y) = \sup_{v \in V} F(e_1(y), \ldots, e_L(y), v)(y)$ is also symmetric.
\end{proof}

	\subsection{Proof of Lemma  \ref{optimality}}\label{sec:pflemma2}

\begin{proof}
	We show that for any synthetic e-merging function $G$, there exists a convex synthetic e-merging function $H$ such that $H$ takes the value $1/\alpha$ on the support of $G$. Let 
$ \{ X^{(1)}, \dots, X^{(n)} \}$ be the support of $G$. Write  
$\ X^{(j)} = (X_1^{(j)}, \dots, X_L^{(j)})^\top$.
	Consider the following system of inequalities:
	\[
	\lambda_1 X_1^{(j)} + \cdots + \lambda_L X_L^{(j)} \geq \frac{1}{\alpha}, \quad j = 1, \dots, n,
	\]
	\[
	\sum_{i=1}^{L} \lambda_i = 1, \quad \lambda_i \geq 0, \quad i = 1, \dots, L.
	\]
	If there exists $\lambda_1,\ldots,\lambda_L$ that satisfies the above system of inequalities then we can define \( H(x) = \frac{1}{\alpha} \ind \left( \lambda_1 x_1 + \cdots + \lambda_L x_L \geq \frac{1}{\alpha} \right) \) and $H(x)=1/ \alpha$ whenever $G(x)=1/\alpha$. Hence $H\geq G$. In what follows we show that there indeed exists  $\lambda_1,\ldots,\lambda_L$ that satisfies the above system.

	Since $  \lambda_L =1 - \sum_{i=1}^{L-1} \lambda_i \geq 0$, the above system can be written equivalently as
	\[
	\lambda_1 (X_1^{(j)} - X_L^{(j)}) + \cdots + \lambda_{L-1} (X_{L-1}^{(j)} - X_L^{(j)}) \geq \frac{1}{\alpha} - X_L^{(j)}, \quad j = 1, \dots, n,
	\]
	\[
	\sum_{i=1}^{L-1} \lambda_i \leq 1, \quad \lambda_i \geq 0, \quad i=1, \dots, L-1.
	\]
	
	Now, we define the matrix \( A \), the vector \( X \), and the vector \( b \) as follows:
	\[
	A = \begin{pmatrix}
		X_1^{(1)} - X_L^{(1)} & \cdots & X_{L-1}^{(1)} - X_L^{(1)} \\
		\vdots & \ddots & \vdots \\
		X_1^{(n)} - X_L^{(n)} & \cdots & X_{L-1}^{(n)} - X_L^{(n)} \\
		-1 & \cdots & -1
	\end{pmatrix},\quad 
	\Lambda = \begin{pmatrix} \lambda_1 \\ \vdots \\ \lambda_{L-1} \end{pmatrix}, \quad
	b = \begin{pmatrix}
		\frac{1}{\alpha} - X_L^{(1)} \\
		\vdots \\
		\frac{1}{\alpha} - X_L^{(n)} \\
		-1
	\end{pmatrix}
	\]
	
	The system can then be written as
	\begin{equation}\label{sys}
		A \Lambda \geq b, \quad \Lambda \geq 0.
	\end{equation}

	We need the following lemma.
	\begin{lemma}[Farkas' Lemma]\cite{Farkas2006}\label{Farkas}
		Either the system of linear inequalities
		\[
		A x \geq b \quad \text{for some } x \geq 0
		\]
		has a solution, or the system
		\[
		A^\top y \leq 0 \quad \text{for some } y \geq 0
		\]
		has a solution with $
		b^\top y > 0.
		$
		
	\end{lemma}
	By Lemma \ref{Farkas}, if the system in \eqref{sys} has no solution then 
	there exists a vector \( y \geq 0 \) such that:
	\[
	A^\top y \leq 0, \quad b^\top y > 0 .
	\]
	Write \( y^\top = (y_1,\ldots, y_{n+1})\) 
	This leads to the following conditions:
	\[
	\sum_{j=1}^{n} (X_i^{(j)} - X_L^{(j)}) y_j - y_{n+1} \leq 0, \quad i = 1, \dots, L-1,
	\]
	and
	\[
	\sum_{j=1}^{n} \left( \frac{1}{\alpha} - X_L^{(j)} \right) y_j - y_{n+1} > 0.
	\]
	
	Now, assume \( y \neq 0 \), as otherwise \( b^\top y = 0 \). Denote
	$
	P_j := \frac{y_j}{\sum_{z=1}^{n+1} y_z}.
	$
	This leads to the transformed system of inequalities:
	\[
	\sum_{j=1}^{n} (X_i^{(j)} - X_L^{(j)}) P_j - P_{n+1} \leq 0, \quad i = 1, \dots, L-1,
	\]
	
	and
	\[
	\sum_{j=1}^{n} \left( \frac{1}{\alpha} - X_L^{(j)} \right) P_j - P_{n+1} > 0.
	\]
	
	To simplify notation, let:
	\[
	U_i := \sum_{j=1}^{n} X_i^{(j)} P_j.
	\]
	The system can then be written as:
\begin{equation}\label{pf1}
	U_i - U_L - P_{n+1} \leq 0, \quad i = 1, \dots, L-1,\quad  \frac{1}{\alpha}(1 - P_{n+1}) - U_L - P_{n+1} > 0.
\end{equation}

	We consider the following two cases.

	\textbf{Case 1: \( U_L + P_{n+1} \geq 1 - P_{n+1} \)}
	
	Define \( e \) as:
	\[
	e = X^{(j)} \text{ with probability } \frac{P_j}{U_L + P_{n+1}},
	\]
	and \( e = (0, \dots, 0) \) with probability \( 1 - \frac{1 - P_{n+1}}{U_L + P_{n+1}} \).
		Since \( U_L + P_{n+1} \geq 1 - P_{n+1} \), we have:
	\[
	\sum_{j=1}^{n} \frac{P_j}{U_L + P_{n+1}} = \frac{1 - P_{n+1}}{U_L + P_{n+1}} \leq 1.
	\]
		Then, by \eqref{pf1} we have 
	\[
	\mathbb{E}(e_i) = \frac{1}{U_L + P_{n+1}} U_i \leq  1, \quad i = 1, \dots, L - 1.
	\]
		For \( e_L \), we have:
	\[
	\mathbb{E}(e_L) = \frac{U_L}{U_L + P_{n+1}} < 1.
	\]
	Thus, each component of \( e \) is an e-value.
		Next, we consider the inequality:
	\[
	\frac{1}{\alpha}(1 - P_{n+1}) - U_L - P_{n+1} > 0 \quad \Longrightarrow \quad \frac{1}{\alpha}(1 - P_{n+1}) > U_L + P_{n+1} \quad \Longrightarrow \quad \frac{1 - P_{n+1}}{U_L + P_{n+1}} > \alpha.
	\]
Since $ \{ X^{(1)}, \dots, X^{(n)} \}$ is the support of $G$ we have
	\[
	\mathbb{P}\left(G(e) = \frac{1}{\alpha}\right) = \sum_{j=1}^{n}\mathbb{P}    (e= X^{(j)}) =\frac{1 - P_{n+1}}{U_L + P_{n+1}} > \alpha.
	\]
	This contradicts the validity of $G$ since  \( \mathbb{E}(G(e)) > 1 \).
	
	\textbf{Case 2: \( U_L + P_{n+1} < 1 - P_{n+1} \)}
	
	Define \( e \) as:
	\[
	e = X^{(j)} \text{ with probability } \frac{P_j}{1 - P_{n+1}}.
	\]
We have
	\[
	\mathbb{E}(e_i) = \frac{1}{1 - P_{n+1}} U_i \leq \frac{1}{1 - P_{n+1}} (U_L + P_{n+1}) = \frac{U_L + P_{n+1}}{1 - P_{n+1}} < 1, \quad i = 1, \dots, L - 1.
	\]
	and
	\[
	\mathbb{E}(e_L) = \frac{1}{1 - P_{n+1}} U_L \leq \frac{U_L + P_{n+1}}{1 - P_{n+1}} < 1.
	\]
Thus, each component of $e$ is an e-value.
	However, we have:
	\[
	\mathbb{P}\left(G(e) = \frac{1}{\alpha}  \right) =
	 \sum_{j=1}^{n}\mathbb{P}(e=X^{(j)})  =1> \alpha.
	\]
	This implies 
	 \( \mathbb{E}(G(e)) > 1 \), a contradiction.
\end{proof}

		  \begin{remark}\label{counterexample}
		A  Convex e-merging function is not necessarily admissible. For example, consider the convex e-merging function \( F \) defined by:
		\[
		F(x) := \frac{1}{\alpha} \ind\left(\frac{1}{L}\sum_{\ell=1}^{L}x_\ell \geq \frac{1}{\alpha}\right)
		\]
		
		For \( L = 2 \), \(\alpha_1 = \alpha_2 \) and \( \alpha_1 \leq \alpha < 2\alpha_1 \),$F$ can be described as follows
		\[
		F: (0, 0), \left( 0, \frac{1}{\alpha_1} \right), \left( \frac{1}{\alpha_1}, 0 \right) \to 0, \quad \left( \frac{1}{\alpha_1}, \frac{1}{\alpha_1} \right) \to \frac{1}{\alpha}.
		\]
		
		Next, consider the function \( G \), which is given by:
		\[
		G: (0, 0), \left( \frac{1}{\alpha_1}, 0 \right) \to 0, \quad \left( 0, \frac{1}{\alpha_1} \right), \left( \frac{1}{\alpha_1}, \frac{1}{\alpha_1} \right) \to \frac{1}{\alpha}.
		\]
		
	Note that $G\geq F$ with strict inequality on the input $(0,1/\alpha_1)$. Next we show that $G$ is valid. Suppose $e=(e_1,e_2)$ where $e_\ell \in \{0,1/\alpha_\ell\}$ and $\EE(e_\ell)\leq 1$.
		We have the following inequality:
		\[
		\mathbb{P}\left(G(e) = \frac{1}{\alpha}\right) \leq \mathbb{P}\left(e_2 = \frac{1}{\alpha_1}\right) \leq \alpha_1 \leq \alpha.
		\]
Consequently,
	$
		\mathbb{E}(G(e)) \leq 1.
		$
			Thus, \( G \) is a valid synthetic e-merging function, which implies $F$ is not admissible.
	\end{remark}

			\subsection{Proof of Lemma \ref{symmetric}}
\begin{proof}
	Note that in this case, the output of a symmetric synthetic e-merging function is determined by the number of nonzero components in the input vector.
	Observe that if $G$ is a valid symmetric synthetic e-merging function then it must map the zero vector to 0. This is because each component of the zero vector is 0 therefore trivially an e-value. But the image of the zero vector is $1/\alpha >1$ a contradiction.  
	 Consequently, we can represent it using a an $L$-bit binary number. More precisely, if the output of $G$ is equal to $1/\alpha$ only when the input vectors have  $n_1,\ldots, n_r$ non-zero entries, we represent it as 0,...,0,1,0,..0,1,..0 where only the $n_1,\ldots, n_r$th positions are 1 and the rest are 0.
	
	
	
	We divide the proof into several cases:
	
	Case $i$: \( (i-1) \alpha < L \alpha_1 \leq i \alpha \ \text{for} \ i = 1,...,L, \)

	Case \( L+1 \): \(\alpha_1 > \alpha \).
	
	For case \( i \leq L \), note that $ F(x):= \frac{1}{\alpha} \ind\left(\frac{1}{L}\sum_{\ell=1}^{L}x_\ell\geq \frac{1}{\alpha}\right)$ can be represented as \( F = 0 \dots 01 \dots 1 \) with \( i-1 \) zeros and \( L-i+1 \) ones. We show that a valid symmetric synthetic e-merging function can only take the form \( 0 \cdots 0 XX \cdots X  \) with the first $i-1$ positions all being 0. This implies $F$ is the only admissible function.
	
	For case $L+1$, We claim that $F$ is the only valid synthetic e-merging function, and naturally, the only admissible function.
	
	For case \( i \leq L \):
	\[
	(i-1) \alpha < L \alpha_1 \leq i \alpha \quad \Longleftrightarrow \quad \frac{i-1}{L \alpha_1} < \frac{1}{\alpha} \leq \frac{i}{L \alpha_1}.
	\]
	
	Let \( G := 0, \dots, 010, \dots, 0 \), where 1 is at the \( q \)-th position, with \( 1 \leq q \leq i-1 \). We claim that \( G \) is not a valid synthetic e-merging function.
	
	If \( 1 - \frac{L \alpha_1}{q} \geq 0 \), we can define the following set:
	\[
	A_q := \{ y \mid y \text{ has exactly } q \text{ non-zero components} \} = \{ y_1, \dots, y_{\binom{L}{q}} \}.
	\]
	Define a random vector $e$ such that 
	\[
	\mathbb{P}(e = y) = \frac{\alpha_1}{\binom{L-1}{q-1}}, \quad \forall y \in A_q,
	\]
	and the remaining probability is allocated to the zero vector.
	
	The condition \( 1 - \frac{L \alpha_1}{q} \geq 0 \) ensures that 
	\[
	\sum_{y \in A_q} \mathbb{P}(e = y) = \binom{L}{q} \cdot \frac{\alpha_1}{\binom{L-1}{q-1}} = \frac{L \alpha_1}{q} \leq 1.
	\]
	
	We now prove that each component of $e$ is an e-value. Taking $e_1$ as an example:
	\begin{align*}
		\mathbb{P}\left(e_1 = \frac{1}{\alpha_1}\right) &= \sum_{x_2, \dots, x_L} \mathbb{P}\left(e_1 = \frac{1}{\alpha_1}, e_2 = x_2, \dots, e_L = x_L\right) \\
		&= \sum_{x_2, \dots, x_L \text{ with exactly } q-1 \text{ non-zeros}} \frac{\alpha_1}{\binom{L-1}{q-1}} \\
		&= \binom{L-1}{q-1} \cdot \frac{\alpha_1}{\binom{L-1}{q-1}} \\
		&= \alpha_1.
	\end{align*}
	
	The same applies for the other \( e_\ell \)'s.	However, we have:
	\begin{align*}
		\mathbb{P}\left( G(e) = \frac{1}{\alpha}\right) &= \binom{L}{q} \cdot \frac{\alpha_1}{\binom{L-1}{q-1}} \\
		&= \frac{L \alpha_1}{q} \\
		&\geq \frac{L \alpha_1}{i-1} > \alpha.
	\end{align*}
	This contradicts the validity of $G$.	If \( 1 - \frac{L \alpha_1}{q} = 1 - \binom{L}{q} \cdot \frac{\alpha_1}{\binom{L-1}{q-1}} < 0 \), then there exists a \( t < \binom{L}{q} \in \mathbb{N} \) satisfying:
	\[
	\frac{t \alpha_1}{\binom{L-1}{q-1}} \leq 1, \quad \frac{(t+1) \alpha_1}{\binom{L-1}{q-1}} > 1.
	\]
	
	Next, we define an $e$ as follows:
	\[
	\mathbb{P}(e = y_s) = \frac{\alpha_1}{\binom{L-1}{q-1}}, \quad s = 1, \dots, t.
	\]
	\[
	\mathbb{P}(e = y_{t+1}) = 1 - \frac{t \alpha_1}{\binom{L-1}{q-1}} < \frac{\alpha_1}{\binom{L-1}{q-1}}.
	\]
	
	We claim that each component of $e$ is an e-value. Taking \( e_1 \) as an example:
	\begin{align*}
		\mathbb{P}\left(e_1 = \frac{1}{\alpha_1}\right) &= \sum_{x_2, \dots, x_L} \mathbb{P}\left(e_1 = \frac{1}{\alpha_1}, e_2 = x_2, \dots, e_L = x_L\right) \\
		&= \sum_{x_2, \dots, x_L \text{ with exactly } q-1 \text{ non-zeros}} \mathbb{P}\left(e_1 = \frac{1}{\alpha_1}, e_2 = x_2, \dots, e_L = x_L\right) \\
		&\leq \sum_{x_2, \dots, x_L \text{ with exactly } q-1 \text{ non-zeros}} \frac{\alpha_1}{\binom{L-1}{q-1}} \\
		&= \binom{L-1}{q-1} \cdot \frac{\alpha_1}{\binom{L-1}{q-1}} \\
		&= \alpha_1.
	\end{align*}
	Similar as before, we have
	\[
	\mathbb{P}\left(G(e) = \frac{1}{\alpha}\right) = \frac{t \alpha_1}{\binom{L-1}{q-1}} +  1 - \frac{t \alpha_1}{\binom{L-1}{q-1}} = 1 > \alpha.
	\]
	This again contradicts the validity of G. Consequently,
	for any \( h \) satisfying \( G \leq h \), \( h \) is also not a valid  e-merging function. Therefore, a valid symmetric synthetic e-merging function can only take the form \( 0 \cdots 0 XX \cdots X  \) with the first $i-1$ positions all being 0, which means $F$ is the only admissible e-merging function.

	For case \( L+1 \)	we claim that \( F(x)\equiv 0 \) is the only valid symmetric e-merging function: 
	
	If there exists symmetric synthetic e-merging function \( G \) and \( x_0  \) s.t. \( G(x_0) = \frac{1}{\alpha} \), assume \( x_0 \) has \( k \) non-zero components. Consider the following $e$
	\[
	e := 
	\begin{cases} 
		\left( \frac{1}{\alpha_1}, \frac{1}{\alpha_1}, \dots, \frac{1}{\alpha_1}, 0, 0, \dots, 0 \right) & \text{the first $k$ entries being $1/\alpha_1$ with probability } \alpha_1, \\
		\left( 0, 0, \dots, 0 \right) & \text{with probability } 1 - \alpha_1.
	\end{cases}
	\]
	
	\[
	\mathbb{P}\left(G(e) = \frac{1}{\alpha}\right) = \alpha_1 > \alpha.
	\]
	Each component of $e$ is an e-value but $\mathbb{E}(G(e)) >1$ a contradiction.
\end{proof}

		\section{Examples of Calibrators in Theorem} \label{ap:discuss_thm_dep_power}
		This section provides examples of calibrators appearing in Theorem \ref{thm:dependent_power}.
	
	In this section, we provide two examples of calibrators: arithmetic mean and R\"uger's method.
	It is easy to check that they both satisfy the assumptions of Theorem \ref{thm:dependent_power}.
	Next, we show that the set $\{y\in\cY: \Gamma(y,\alpha,\alpha') > 1\}$ is not empty for these two calibrators across a wide range of scenarios.
	
	\noindent\textbf{Arithmetic Mean.}
		We first claim that for any $0<\alpha<\alpha'<1$, if $f_\ell$ is defined by the calibrator of arithmetic mean $f(p) = (2-2p)\ind\{p\in(0,1)\} + \infty\ind(p=0)$ in \eqref{eq:combine_dep_general}, we have $\Gamma(y,\alpha,\alpha') > 1$ 
		if and only if
		\begin{align*}
			\PP(y\notin \cC_{1,\alpha}) > (1-\alpha)\cdot\frac{(1-\alpha')-(1-\alpha')^2}{(1-\alpha)-(1-\alpha')^2}+\alpha.
		\end{align*}
		
		To prove this claim, note that by elementary calculations, it holds that
		\begin{align}
			\frac{\alpha'}{\alpha}\int_0^{\alpha/\alpha'}f\left(t\right)\rd t-\frac{\alpha'}{1-\alpha}\int_{\alpha/\alpha'}^1f\left(t\right)\rd t
			=\frac{(1-\alpha)-(1-\alpha')^2}{\alpha'(1-\alpha)}>0.
			\label{eq:eg_ar_s}
		\end{align}
		Recall that by Theorem \ref{thm:dependent_power}, we have 
		\begin{align*}
			\Gamma(\theta,\alpha,\alpha')=\left(\frac{\alpha'}{\alpha}\int_0^{\alpha/\alpha'}f\left(t\right)\rd t-\frac{\alpha'}{1-\alpha}\int_{\alpha/\alpha'}^1f\left(t\right)\rd t\right)\cdot\left(\PP(y\notin \cC_{1,\alpha})-\alpha\right)+\alpha'.
		\end{align*}
		Therefore, using \eqref{eq:eg_ar_s} we can rewrite $\Gamma(\theta,\alpha,\alpha') > 1$ as
		\begin{align*}
			&\PP(y\notin \cC_{1,\alpha}) > \frac{\alpha'(1-\alpha')(1-\alpha)}{(1-\alpha)-(1-\alpha')^2} + \alpha=(1-\alpha)\frac{(1-\alpha')-(1-\alpha')^2}{(1-\alpha)-(1-\alpha')^2} + \alpha.
		\end{align*}
		Then by the fact that
		\[
		(1-\alpha)\frac{(1-\alpha')-(1-\alpha')^2}{(1-\alpha)-(1-\alpha')^2} + \alpha < (1-\alpha)\frac{1-\alpha'}{1-\alpha} + \alpha = 1-\alpha'+\alpha,
		\]
		we show that $\{y\in\cY: \Gamma(y,\alpha,\alpha') > 1\}$ is not empty as long as $0<\alpha<\alpha'<1$.

	\noindent\textbf{R\"uger's Method.}
		We first claim that for any $0<\alpha<\alpha'<1$, if $f_\ell$ is defined by R\"uger's method, i.e., $f(p)=L/k\cdot\ind\{p\in(0,k/L)\}+\infty \ind(p=0)$, then $\Gamma(y,\alpha,\alpha') > 1$ if and only if
		\begin{align}
			\PP(y\notin \cC_{1,\alpha}) > \alpha(1-\alpha')\left(\alpha'-C_k\frac{\alpha}{1-\alpha}\right)^{-1}+\alpha,
			\label{eq:cal_rug}
		\end{align}
		where $C_k = \left(\alpha'/\alpha - L/k\right)_+ \in [0, \alpha'/\alpha-1]$. 
		To verify this, we consider two cases.
		
		(a). If $\alpha/\alpha' < k/L$, we have 
			\begin{align}
				\frac{\alpha'}{\alpha}\int_0^{\alpha/\alpha'}f\left(t\right)\rd t-\frac{\alpha'}{1-\alpha}\int_{\alpha/\alpha'}^1f\left(t\right)\rd t
				=\frac{1}{1-\alpha}\left(\frac{L}{k}-\alpha'\right)>0.
				\label{eq:eg_rk_s1}
			\end{align}
			Combining \eqref{eq:eg_rk_s1} and the definition of $\Gamma(\theta,\alpha,\alpha')$ in Theorem \ref{thm:dependent_power}, by plugging in $C_k = \alpha'/\alpha - L/k$, we have that $\Gamma(y,\alpha,\alpha') > 1$ is equivalent to 
			\begin{align}
				&\PP(y\notin \cC_{1,\alpha}) > \frac{(1-\alpha')(1-\alpha)}{\alpha'/\alpha - C_k-\alpha'}+\alpha=\frac{\alpha(1-\alpha')}{\alpha'-C_k\frac{\alpha}{1-\alpha}}+\alpha.
				\label{eq:eg_rk_res1}
			\end{align}
			
			(b). If $\alpha/\alpha' \geq k/L$, then we have $C_k = 0$ and
			$
				\frac{\alpha'}{\alpha}\int_0^{\alpha/\alpha'}f\left(t\right)\rd t-\frac{\alpha'}{1-\alpha}\int_{\alpha/\alpha'}^1f\left(t\right)\rd t = \frac{\alpha'}{\alpha}\frac{k}{L}\frac{L}{k} = \frac{\alpha'}{\alpha}.
			$
			It follows that $\Gamma(y,\alpha,\alpha') > 1$ holds if and only if
			\begin{align}
				\PP(y\notin \cC_{1,\alpha}) > (1-\alpha')\frac{\alpha}{\alpha'}+\alpha = \frac{\alpha}{\alpha'}.
				\label{eq:eg_rk_res2}
			\end{align}
			Combining \eqref{eq:eg_rk_res1} and \eqref{eq:eg_rk_res2}, the statement in \eqref{eq:cal_rug} holds for any $k\in[L]$. 
			Therefore, the set $\{y\in\cY: \Gamma(y,\alpha,\alpha') > 1\}$ is not empty as long as $0<\alpha<\alpha'<1$ and $k \in [L-1]$.

		\section{Sensitivity Analysis on Synthetic p-values}\label{ap:sensitivity}
\begin{figure}[ht]
			\centering
			\includegraphics[width=0.95\linewidth]{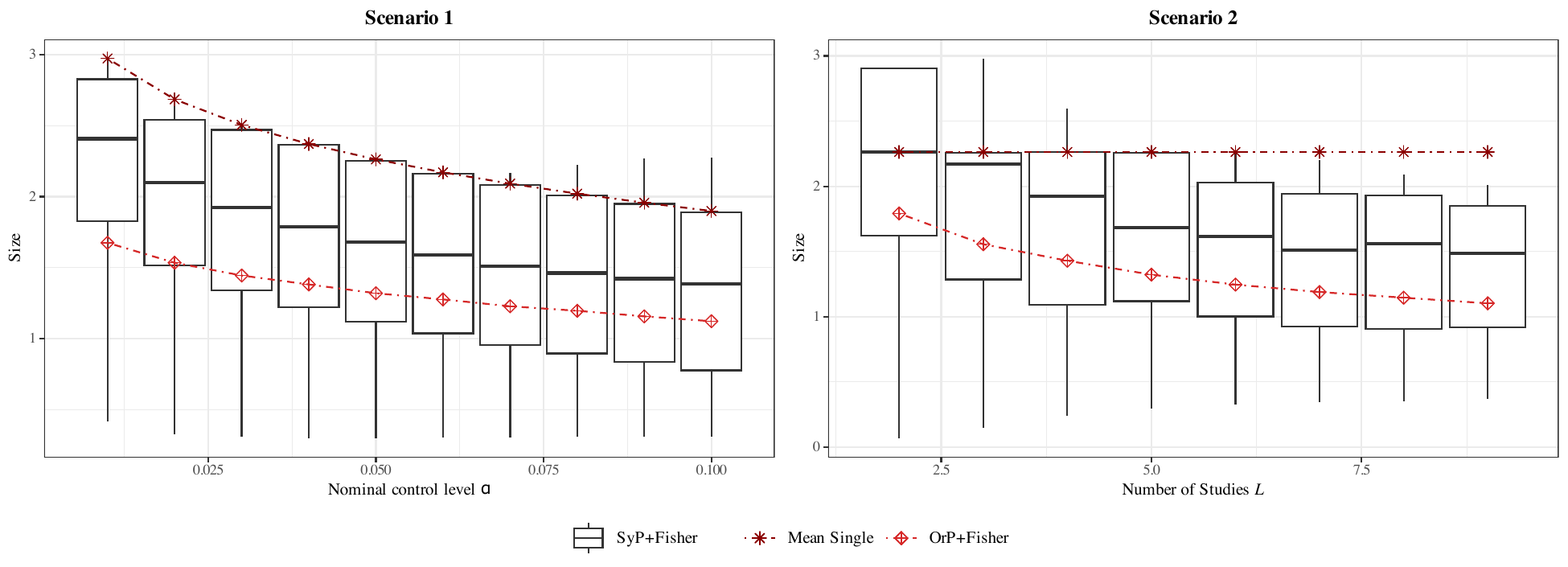}
			\caption[Sensitivity analysis on synthetic p-values]{Sensitivity analysis on synthetic p-values for Scenarios 1 and 2 in Section \ref{sec:simu1}.}
			\label{fig:sim_sensitivity}
		\end{figure}

		As synthetic p-values are random, a natural question arises: how sensitive is the resulting merged set to this randomization? 
		This section is dedicated to exploring this sensitivity using the framework outlined in Section \ref{sec:simu1}. 
		Specifically, within each replication of Scenarios 1 and 2, we independently run \texttt{SyP+Fisher} $2000$ times and record the minimum, maximum, median, and the upper and lower 2.5\% quantiles of the sizes of the final merged sets. We then replicate this process $5000$ times and calculate the mean for each statistic.  
		Finally, we present the results via box plots, comparing the sizes of the merged sets produced by \texttt{SyP+Fisher} against those generated by an individual study and by \texttt{OrP+Fisher} in Figure \ref{fig:sim_sensitivity}. 
		Despite the randomness introduced by synthetic p-values, we observe that \texttt{SyP+Fisher} consistently reduces the sizes of uncertainty sets compared with the set size of the individual study.  
		The median sizes of \texttt{SyP+Fisher} are close to those of \texttt{OrP+Fisher}, indicating effective utilization of information from the initial uncertainty sets. 
		Furthermore, as $L$ increases, the sizes of \texttt{SyP+Fisher} tend to decrease, demonstrating that the proposed procedure can leverage more information from an expanding number of initial uncertainty sets.
	
\section{Admissible Deterministic Set Merging under Independence}\label{ap:merge_indep_no_random}
As discussed in the proof of Theorem \ref{admissibility}, merging uncertainty sets is equivalent to merging e-functions. By Lemma \ref{lemma1}, this further reduces to studying local e-function mergers.
A function $F: \prod_{\ell=1}^{L} \{0, 1/\alpha_\ell\} \to \{0, 1/\alpha\}$ is a synthetic independent e-value (ie) merging function if it preserves the e-value property under independence: for any vector of independent random variables $\pmb{e}=(e_1, \ldots, e_L)$ where each component $e_\ell$ satisfies $\EE(e_\ell)\le 1$, the output $F(\pmb{e})$ must also satisfy $\EE\{F(\pmb{e})\} \le 1$.
A synthetic ie-merging function $F$ is called admissible if there does not exist another synthetic ie-merging function $G$ such that $G \geq F$ and $G\neq F$. We have the following sufficient condition for the admissibility of synthetic ie-merging function:

\begin{proposition}\label{prop:ieadmissible} For any non-decreasing function $F: \{0, 1/\alpha_l\}_{l=1}^{L} \to \{0, 1/\alpha\}$, if it satisfies that for any synthetic ie-vector $\pmb{e}$ with $\mathbb{E}(\pmb{e}) = (1, \dots, 1)$, we have $\mathbb{E}\{F(\pmb{e})\} = 1$, then $F$ is an admissible synthetic ie-merging function.
\end{proposition}
\begin{proof} 
	The proof proceeds in two parts. First, we show that any function $F$ satisfying the condition is a valid synthetic ie-merging function. Second, we show by contradiction that it must be admissible.
	
	(i) \textit{Validity}. Let $\pmb{e}=(e_1, \ldots, e_L)$ be a vector of independent synthetic e-values, meaning each $e_\ell$ is supported on $\{0, 1/\alpha_\ell\}$ and satisfies $\EE(e_\ell) \le 1$. Let $p_\ell = \PP(e_\ell = 1/\alpha_\ell)$, which implies $p_\ell \le \alpha_\ell$. We construct a new vector of independent random variables $\tilde{\pmb{e}}=(\tilde{e}_1, \ldots, \tilde{e}_L)$ such that $\EE(\tilde{\pmb{e}}) = \mathbf{1}$. For each $\ell=1, \ldots, L$, we define $\tilde{e}_\ell$ based on the outcome of $e_\ell$:
	\begin{itemize}
		\item if $e_\ell = 1/\alpha_\ell$, we set $\tilde{e}_\ell = 1/\alpha_\ell$.
		\item if $e_\ell = 0$, we draw $\tilde{e}_\ell$ from a distribution such that $\PP(\tilde{e}_\ell = 1/\alpha_\ell) = (\alpha_\ell - p_\ell)/(1-p_\ell)$ and $\PP(\tilde{e}_\ell = 0)$ is the remaining probability.
	\end{itemize}
	By construction, $\tilde{e}_\ell \ge e_\ell$ for all outcomes. The unconditional expectation of $\tilde{e}_\ell$ is
	\[
	\EE(\tilde{e}_\ell) = \PP(e_\ell = 1/\alpha_\ell) \cdot \frac{1}{\alpha_\ell} + \PP(e_\ell = 0) \cdot \left( \frac{1}{\alpha_\ell} \cdot \frac{\alpha_\ell - p_\ell}{1-p_\ell} \right) = \frac{p_\ell}{\alpha_\ell} + \frac{\alpha_\ell - p_\ell}{\alpha_\ell} = 1.
	\]
	Since the components of $\tilde{\pmb{e}}$ are independent and have an expectation of 1, the premise of the proposition applies, and we have $\EE\{F(\tilde{\pmb{e}})\} = 1$. Because $F$ is a non-decreasing function and $\tilde{\pmb{e}} \ge \pmb{e}$ component-wise, it follows that $F(\tilde{\pmb{e}}) \ge F(\pmb{e})$. Therefore, $\EE\{F(\pmb{e})\} \le \EE\{F(\tilde{\pmb{e}})\} = 1$, which confirms that $F$ is a valid synthetic ie-merging function.
	
	(ii) \textit{Admissibility}. We now prove admissibility by contradiction. Assume $F$ is not admissible. Then there must exist another valid ie-merging function $G$ such that $G(\pmb{e}) \ge F(\pmb{e})$ for all $\pmb{e}$, and $G(\pmb{e}^*) > F(\pmb{e}^*)$ for at least one specific outcome $\pmb{e}^* \in \prod_{\ell=1}^{L} \{0, 1/\alpha_\ell\}$.
	
	To construct the contradiction, consider a specific vector of independent random variables, $\pmb{e}^\dagger = (e^\dagger_1, \ldots, e^\dagger_L)$, where each component $e^\dagger_\ell$ is a Bernoulli trial scaled by $1/\alpha_\ell$: it takes the value $1/\alpha_\ell$ with probability $\alpha_\ell$ and $0$ with probability $1-\alpha_\ell$. By construction, $\EE(e^\dagger_\ell) = 1$ for all $\ell$, so $\EE(\pmb{e}^\dagger) = \mathbf{1}$.
	
	By the premise of the proposition, $\EE\{F(\pmb{e}^\dagger)\} = 1$. However, because $G(\pmb{e}) \ge F(\pmb{e})$ for all outcomes, and is strictly greater for the outcome $\pmb{e}^*$ which occurs with a non-zero probability $\PP(\pmb{e}^\dagger = \pmb{e}^*) > 0$, the expectation of $G$ must be strictly greater than the expectation of $F$:
	\[
	\EE\{G(\pmb{e}^\dagger)\} > \EE\{F(\pmb{e}^\dagger)\} = 1.
	\]
	This contradicts the fact that $G$ is a valid ie-merging function, as its expectation exceeds 1. Therefore, our assumption was false, and $F$ must be admissible.
\end{proof}

Any function $F: \prod_{\ell=1}^{L} \{0, 1/\alpha_\ell\} \to \{0, 1/\alpha\}$ can be expressed in the form $F(\pmb{e}) = \alpha^{-1} \ind\{f(\pmb{e}) \in A\}$ for some function $f$ and a set $A$. 
Let $\pmb{e}^\dagger=(e^\dagger_1, \ldots, e^\dagger_L)$ be a random vector whose components are independent and follows $\frac{1}{\alpha_\ell}\text{Ber}(\alpha_\ell)$ distribution. 
Since we are considering synthetic ie-merging function the input of $F$ must have the same distribution as  $\pmb{e}^\dagger$.
For such $\pmb{e}^\dagger$, the condition $\EE\{F(\pmb{e}^\dagger)\} = 1$ from Proposition \ref{prop:ieadmissible} is equivalent to requiring that $\PP\{f(\pmb{e}^\dagger) \in A\} = \alpha$.

However, this equality is often impossible to satisfy exactly. The random variables $e^\dagger_\ell$ are discrete, so the probability $\PP\{f(\pmb{e}^\dagger) \in A\}$ is a sum of probabilities of elementary outcomes. The sample space consists of $2^L$ atoms, and the probability of an outcome where the non-zero indices are in a set $S \subseteq \{1, \ldots, L\}$ is $\prod_{\ell \in S} \alpha_\ell \prod_{\ell \notin S} (1 - \alpha_\ell)$. The set of all possible probabilities for $\PP\{f(\pmb{e}^\dagger) \in A\}$ is therefore a finite set of sums of these values. If a given target level $\alpha$ does not belong to this discrete set of attainable probabilities, the sufficient condition cannot be met.

This motivates the need for a more complete, necessary and sufficient characterization of admissibility, which we provide in the following proposition.

\begin{proposition} \label{prop:iff_admissible}
A synthetic ie-merging function $F(\pmb{e}) = \alpha^{-1} \ind\{f(\pmb{e}) \in A\}$ is admissible if and only if there is no outcome $\pmb{b}$ in the zero set of $F$, $B = \{ \pmb{e} \mid F(\pmb{e}) = 0 \}$, such that
\begin{equation}\label{eq:ieadmissible}
	\PP(\pmb{e}^\dagger = \pmb{b}) \leq \alpha - \PP\{F(\pmb{e}^\dagger) = 1/\alpha\},
\end{equation}
where $\pmb{e}^\dagger$ is the random vector whose components are independent and follows $\frac{1}{\alpha_\ell}\text{Ber}(\alpha_\ell)$ distribution. 
\end{proposition}

\begin{proof}
	($\Rightarrow$) Suppose for contradiction that an admissible function $F$ violates the condition, i.e., there exists a $\pmb{b} \in B$ for which \eqref{eq:ieadmissible} holds. Define a new function $G$ such that $G(\pmb{e}) = 1/\alpha$ if $F(\pmb{e}) = 1/\alpha$ or if $\pmb{e} = \pmb{b}$, and $G(\pmb{e}) = 0$ otherwise. By construction, $G \ge F$ and $G \neq F$. The expectation of $G$ under the distribution of $\pmb{e}^\dagger$ is
	\[
	\EE\{G(\pmb{e}^\dagger)\} = \frac{1}{\alpha} \left[ \PP\{F(\pmb{e}^\dagger) = 1/\alpha\} + \PP(\pmb{e}^\dagger = \pmb{b}) \right] \le \frac{1}{\alpha} \cdot \alpha = 1.
	\]
	This implies that $G$ is a valid synthetic ie-merging function, which contradicts the admissibility of $F$.
	
	($\Leftarrow$) Conversely, suppose $F$ is not admissible. Then there exists a valid ie-merging function $G$ such that $G \ge F$ and $G \neq F$. This means there must be at least one outcome $\pmb{b}$ in the zero set $B$ of $F$ for which $G(\pmb{b}) = 1/\alpha$. Since $G$ is a valid ie-merging function, $\EE\{G(\pmb{e}^\dagger)\} \le 1$, which implies $\PP\{G(\pmb{e}^\dagger) = 1/\alpha\} \le \alpha$. Furthermore, because $G$ dominates $F$ and is strictly greater at $\pmb{b}$,
	\[
	\PP\{G(\pmb{e}^\dagger) = 1/\alpha\} \ge \PP\{F(\pmb{e}^\dagger) = 1/\alpha\} + \PP(\pmb{e}^\dagger = \pmb{b}).
	\]
	Combining these inequalities yields $\PP(\pmb{e}^\dagger = \pmb{b}) \le \alpha - \PP\{F(\pmb{e}^\dagger) = 1/\alpha\}$, which violates the condition of the proposition.
\end{proof}

The condition in \eqref{eq:ieadmissible} is verifiable in principle. The sample space of $\pmb{e}^\dagger$ is finite with $2^L$ atoms, and the probability of any specific outcome $\pmb{x} \in \prod \{0, 1/\alpha_\ell\}$ is explicitly computable as $\prod_{\{\ell:\,x_\ell=1/\alpha_\ell\}}\alpha_\ell \prod_{\{\ell:\,x_\ell=0\}}(1-\alpha_\ell)$. By enumerating the atoms in the acceptance set of $f$, we can calculate $\PP\{F(\pmb{e}^\dagger) = 1/\alpha\}$. Thus, the admissibility of any given function $F$ can be determined by checking the inequality for each of the finite number of points in its zero set.
\end{document}